\documentclass[a4paper,fleqn,usenatbib]{mnras}

\usepackage{txfonts}

\usepackage[T1]{fontenc}
\usepackage{ae,aecompl}

\usepackage{graphicx}	

\title[Monte Carlo production uncertainties for $p$ nuclei]{Uncertainties in the production of
 $p$ nuclei in massive stars obtained from Monte Carlo variations}

\author[T. Rauscher et al.]{T. Rauscher,$^{1,2,3}$\thanks{E-mail: Thomas.Rauscher@unibas.ch}
N. Nishimura,$^{3,4}$
R. Hirschi,$^{3,4,5}$
G. Cescutti,$^{2,3}$
\newauthor
A. St. J. Murphy$^{3,6}$
and A. Heger$^{7}$
\\
$^{1}$Department of Physics, University of Basel, Klingelbergstr.\ 82, 4056 Basel, Switzerland\\
$^{2}$Centre for Astrophysics Research, University of Hertfordshire, College Lane, Hatfield AL10 9AB, UK\\
$^{3}$UK Network for Bridging Disciplines of Galactic Chemical Evolution (BRIDGCE), \url{http://www.bridgce.net/}, UK\\
$^{4}$Astrophysics group, Lennard-Jones Laboratories, Keele University, ST5 5BG, Staffordshire, UK\\
$^{5}$Kavli Institute for the Physics and Mathematics of the Universe (WPI),
University of Tokyo, 5-1-5 Kashiwanoha, Kashiwa, 277-8583, Japan\\
$^{6}$SUPA, School of Physics and Astronomy, University of Edinburgh, Edinburgh EH9 3FD, UK\\
$^{7}$School of Physics and Astronomy, Clayton Campus, Monash University, Victoria 3800, Australia
}

\date{Accepted XXX. Received YYY; in original form ZZZ}

\pubyear{2016}

\begin{document}
\label{firstpage}
\pagerange{\pageref{firstpage}--\pageref{lastpage}}
\maketitle

\begin{abstract}
Nuclear data uncertainties in the production of $p$ nuclei in massive stars have been quantified in a Monte Carlo procedure. Bespoke temperature-dependent uncertainties were assigned to different types of reactions involving nuclei from Fe to Bi. Their simultaneous impact was studied in postprocessing explosive trajectories for three different stellar models. It was found that the grid of mass zones in the model of a 25 $M_\odot$ star, which is widely used for investigations of $p$ nucleosynthesis, is too crude to properly resolve the detailed temperature changes required for describing the production of $p$ nuclei. Using models with finer grids for 15 $M_\odot$ and 25 $M_\odot$ stars with initial solar metallicity, it was found that most of the production uncertainties introduced by nuclear reaction uncertainties are smaller than a factor of two. Since a large number of rates were varied at the same time in the Monte Carlo procedure, possible cancellation effects of several uncertainties could be taken into account. Key rates were identified for each $p$ nucleus, which provide the dominant contribution to the production uncertainty. These key rates were found by examining correlations between rate variations and resulting abundance changes. This method is superior to studying flow patterns, especially when the flows are complex, and to individual, sequential variation of a few rates.
\end{abstract}




\section{Introduction}
\label{sec:intro}

It is well established that the astrophysical origin of the majority of nuclides beyond Fe requires at least two neutron-capture processes, the $s$ process and the $r$ process \citep{cam,burb}. The abundances of a comparatively small number of nuclides, however, cannot be explained by those processes.  While older literature counted 35 nuclides, called the $p$ nuclei, being problematic in the context of neutron-capture processes, more recent work also suggests strong s-process contributions to $^{164}$Er, $^{152}$Gd, and $^{180}$Ta \citep{arl99}, and combined s- and r-process contributions to $^{113}$In and $^{115}$Sn \citep{nemeth}. The nucleus $^{138}$La is produced by neutrino-induced reactions in core-collapse supernova explosions of massive stars and also $^{180}$Ta may have some neutrino-induced contributions \citep{neutrino}. This leaves the origin of 30 proton-rich isotopes to be explained.

For detailed explanations of the $p$ nuclides and their possible production in various astrophysical sites, see, e.g., the reviews by \citet{arngor,p-review,pign-rev} and references therein. Here, only a brief account of the most important facts is provided, as relevant to establish the context of our present study.

Solar system $p$ abundances have been derived from geological and meteoritic data. Understanding the origin of the $p$ nuclides is challenging because they cannot be directly observed in stars and supernova remnants, as their contribution to elemental abundances is small and no element is dominated by a $p$ isotope. Therefore, possible nucleosynthesis processes have to be studied in models without the possibility of direct verification. Additionally, terrestrial and meteoritic $p$ abundances have to be derived from Galactic Chemical Evolution (GCE) models, integrating the production of different sites over the history of the Galaxy. Moreover, the solar composition may not follow the average galactic composition, which is calculated in GCE models. Model uncertainties are introduced in each simulation step but the initial uncertainties are due to the astrophysical reaction rates used in the nuclear reaction networks. It is the aim of the current work to quantify the contribution of nuclear uncertainties to the production uncertainties of selected models. It is useful for nucleosynthesis studies to directly see the uncertainties in the abundances of modeled nucleosynthesis processes, as they will guide theoretical and experimental nuclear scientists in improving the knowledge of selected rates, and they can also be included and propagated in GCE models. In order to obtain a quantification of final abundance uncertainties, we apply a Monte Carlo method of varying rates according to their nuclear uncertainties. The method can be generally applied to any nucleosynthesis environment which is studied with reaction networks and also allows identification of key rates in a self-consistent and automatic manner. Here, we apply it to the synthesis of $p$ nuclei but will also study further processes in subsequent papers.

As for the $p$ nuclei, it was suggested that they can be produced by photodisintegration reactions of pre-existing seed nuclei in the outer shells of exploding massive stars \citep{cam,burb,woohow,arn,rayet95,rhhw02}. Although this $\gamma$ process occurs naturally in stellar explosions and thus is able to produce the bulk of $p$ nuclides within a single site, there are longstanding problems in obtaining $p$ abundances consistent with solar system amounts for at least two $p$-nucleus mass ranges: light ones with mass number $A<110$, heavier ones, in the mass range $150\leq A \leq165$ \citep{rhhw02}. These deficiencies in $p$-nucleus production have triggered a large number of investigations in astrophysics and nuclear physics, both theoretically and experimentally. It is commonly assumed that the underproduction in the heavier mass range may be cured by improved nuclear physics input whereas a different astrophysical site may be required for the light $p$ nuclei. A promising alternative site is the explosion of a mass-accreting White Dwarf in a thermonuclear supernova \citep{howmey}. It has been shown that the full range of $p$ nuclei could be produced in such an explosion but a strong production of $s$-process seeds in the accreting matter is required \citep{kusa11,travWD,gorsub3D}. Currently, there are no self-consistent simulations of the accretion, $s$-process enhancement, and explosion, but rather a high level of $s$-process seeds has to be assumed.

In this work, we focus on the $\gamma$ process in massive stars, which depends on the assumed initial abundances and the stellar mass \citep{rhhw02,p-review}. It is also influenced by the modeled stellar evolution and explosion, and thus depends on the stellar model code. To explore both of these dependences, we perform production uncertainty analyses in two different basic stellar models and for two progenitor masses. A similar analysis of the synthesis of $p$-nuclei in a 2D thermonuclear White Dwarf explosion currently is in progress and will be presented in a forthcoming publication.

In addition to the stable $p$ nuclides, the $\gamma$ process also produces unstable isotopes with relatively long half-life \citep[see, e.g.,][]{travradio}. These are important insofar as a number of now extinct radiogenic nuclides were present in the early Solar System. In particular, signatures of the presence of $^{92}$Nb, $^{97,98}$Tc, and $^{146}$Sm have been found, detected as excess of their stable daughter isotopes. For a detailed account of these measurements and their implication for Galactic Chemical Evolution, see, e.g., Chapter 4 of \citet{p-review}. Because of their importance, we also study the production of $^{92}$Nb, $^{146}$Sm and $^{97,98}$Tc in our models and provide the resulting production uncertainties in section \ref{sec:radio}.

\section{Reaction rate variation}
\label{sec:MCmethod}

\subsection{The Monte Carlo approach}
\label{sec:MCgeneral}

The Monte Carlo (MC) method of using random number input to numerical simulations has a long history in science and engineering. It has been used relatively seldomly in astrophysical investigations due to the demand on computation time per run in hydrodynamical simulations \citep[for application to nuclear astrophysics, see, e.g.,][]{ili,longland1,longland2}. With the advent of fast computers and the use of postprocessing techniques it has become feasible, though, to perform a statistically sufficient number of MC iterations even with reaction networks containing several thousand nuclei.

Here, we employ a straightforward application of the MC method by drawing sets of random numbers to simultaneously vary reaction rates within predefined uncertainty limits. This is described in detail in sections \ref{sec:pizbuin} and \ref{sec:uncert}. In this way, the initial nuclear uncertainties are mapped onto uncertainties in the final abundances via the reaction network.

Using such a MC approach allows the final uncertainties to be quantified, even when many individual reactions contribute to the abundance of a given nucleus. While the inspection of reaction flows in flow plots may be sufficient when only few reactions contribute within the range of a well-defined reaction path, the effect of complex flow patterns -- such as those arising in the high-temperature environments of explosive nucleosynthesis -- and the \textit{combined} action of uncertainties can only be addressed using a MC method. The technique also makes possible the identification of those reactions that give rise to the largest uncertainties in individual isotope final abundances, as is presented in Section \ref{sec:correlations}.

\subsection{The PizBuin framework}
\label{sec:pizbuin}

Our MC framework PizBuin consists of an efficient reaction network code and a parallelized MC driver providing random number input with the possibility to choose the number of varied rates and the random distributions to draw from. Another feature is the possibility to define temperature-dependent uncertainty factors for each rate individually. This is further explained in Section \ref{sec:varfacts}. Tools to analyze the MC output complement the code suite.

A nuclear reaction network is a stiff set of coupled differential equations describing the temporal change $\dot{Y}_i=dY_i/dt$ in the abundance of a given nucleus $i$. The network includes all reactions affecting the abundance of a nucleus \citep{rau_ijmpe},
\begin{equation}
\label{eq:network}
 \dot{Y}_i = \frac{1}{\rho N_\mathrm{A}}
 \left\{ \sum_j {^{1}_{i}K_{j}} \; {_{i}\lambda^*_{j}} + \sum_{j} {^{2}_{i}K_{j}}\; {_{i}r^*_{j}} \right\} \quad,
\end{equation}
where $N_\mathrm{A}$ is Avogadro's number, $1\leq i\leq m$ numbers the nucleus, $_i\lambda^*_j$ is the stellar
$j$th rate for destruction or creation of the $i$th nucleus
without a nuclear projectile involved (this includes spontaneous
decay, lepton capture, photodisintegration), and $_ir^*_j$ is the stellar
rate of the $j$th reaction involving a nuclear projectile and creating or
destroying nucleus $i$.  The quantities $^1 _iK_j$ and $^2_iK_j$
are positive or negative integer numbers specifying the number of
nuclei $i$ produced or destroyed, respectively, in the given process. Similarly, three-body reactions can be included but they are not shown here because they were not varied.
As shown below, the stellar rates $\lambda^*$ and $r^*$ contain the abundances of the interacting nuclei and depend on the plasma temperature $T$. Furthermore, the abundance change depends on the density $\rho$ of the plasma. A postprocessing trajectory defines $T$ and $\rho$ as a function of time and thus the above reaction network has to be solved in each timestep when following the evolution of abundances through a nucleosynthesis process. This implies that an $m\times m$ matrix has to be inverted in each timestep. The stiffness of the set of differential equations requires small timesteps. The inversion, however, can be made more efficient due to the fact that this is a highly sparse matrix and sophisticated mathematical techniques can be applied.

The two-body rate $_ir^*_j=r^*_{Aa}$ for a reaction $a+A$ between a projectile $a$ and the nucleus $A$ is given by 
\begin{equation}
r^*_{Aa} =  \frac{Y_A Y_a}{1+\delta_{Aa}} \rho^2 N_A^2 \langle \sigma^*_{Aa} v \rangle \quad, \label{eq:rate}
\end{equation}
with $\langle \sigma^*_{Aa} v \rangle$ being the reactivity, i.e., the stellar cross section (including reactions on thermally populated excited states in $A$) folded with the velocity distribution of the interacting particles.

Rates of two-body reactions including only leptons and photodisintegrations can be expressed similarly to decays in a rate $_i\lambda^*_j=Y_i N_\mathrm{A} L^*_j$. For simple decays, $L^*=(\ln 2) / T^*_{1/2}$ is related to the (stellar) half-life. It has to be noted that $T^*_{1/2}$ depends on the plasma temperature and thus $L^*$ is not always just a constant. For photodisintegrations, $L^* (T)$ is derived from integrating the stellar photodisintegration cross section (including the effect of thermally populated excited states) over a Planck distribution. Alternatively, it can be derived from the capture rate as shown below in equation (\ref{eq:revphoto}). Rates for reactions with leptons (for example, electron captures) can also be written in this form, with $L^* (T,Y_\mathrm{e})$ depending on temperature and electron abundance $Y_\mathrm{e}$.

It is important to remember that forward and reverse rates are connected by the reciprocity relation for nuclear reactions, also called detailed balance. While for weak interactions the inverse processes are not relevant in the current context, temperatures are sufficiently high to allow competition between captures and photodisintegrations as well as between forward and reverse channels of other reactions. The existence of a reciprocity relation for stellar rates is also important when applying multiplicative variation factors because both reaction directions have to be changed by the same factor. This can easily be seen when writing the reverse reactivity (for the reaction $F+b \rightarrow a+A$) in terms of the forward reactivity of the reaction $A+a\rightarrow b+F$,
\begin{equation}
\label{eq:revrate}
\frac{\langle \sigma^*_Fb v\rangle}{\langle \sigma^*_{Aa} v \rangle}=\frac{g_a}{g_b} \frac{G^A}{G^F} \left( \frac{m_{Aa}}{m_{Bb}}\right) ^{3/2}e^{-Q_{Aa}/(k_\mathrm{B}T)} \quad ,
\end{equation}
where $Q_{Aa}$ is the reaction $Q$-value of the forward reaction \citep{rau_ijmpe}. Similarly, photodisintegration $C+\gamma \rightarrow a+A$ is related to capture $A+a\rightarrow \gamma +C$ through
\begin{equation}
\frac{L^*_\gamma}{\langle \sigma^*_{Aa} v \rangle}=g_a \frac{G^A(T)}{G^C(T)}
\left( \frac{m_{Aa}k_\mathrm{B}T}{2\pi \hbar^2}\right)^{3/2} e^{-Q_{Aa}/(kT)} \quad. \label{eq:revphoto}
\end{equation}
These equations make use of the spin factors $g_x=2J_x+1$ of projectile and ejectile, respectively, and of the partition functions $G(T)$ of the participating nuclei, also depending on the plasma temperature $T$. The nuclear partition function $G(T)$ is defined, e.g., in \citet{adndt00}. The reduced mass of projectile and target nucleus is denoted by $m_{Aa}$ and $k_\mathrm{B}$ is the Boltzmann constant.
The above relations only apply to stellar reactivities, not to those derived from cross sections only including ground states of target nuclei. Thus, they do not apply to reactivities derived from experimental cross sections, unless reactions on excited states do not contribute. A criterion for this will be discussed and used in Section \ref{sec:uncert}.

From equations (\ref{eq:revrate}) and (\ref{eq:revphoto}) it is directly seen that multiplying the stellar reactivity $\langle \sigma^*_{Aa} v \rangle$ by a given factor will also increase the reactivity of the reverse reaction by the same factor. Therefore the variation factors derived from the output of the MC driver (see Section \ref{sec:varfacts}) are applied to both reaction directions in each MC iteration and the final abundances calculated by following the trajectories in the reaction network with modified rates. 

The MC driver passes a vector $v_z$ to the network code. This vector contains $z$ random numbers between 0 and 1 for the variation of the chosen $z$ rates to be varied (out of the $n$ reactions). The random values are drawn from preset random distributions. It is possible to choose a different distribution (uniform, normal, lognormal, \dots ) for each rate. When calculating a reaction rate for a trajectory timestep, the network code then maps the random numbers to a variation factor for each rate. Again, the mapping can be different for each rate, depending on preselected choices but also on plasma temperature. Details are given in Section \ref{sec:varfacts}.

Each MC iteration consists of running all available trajectories (see Section \ref{sec:stellarmodels}). At the end of each trajectory run, the final abundances of selected nuclei are recorded. For the present application to the $\gamma$ process, these are the ones of the $p$ nuclei and of a few long-lived radioactivities.
The combined final abundance data are finally analyzed to find the abundance variations, and thus the production uncertainties based on nuclear uncertainties, caused by the variation of the reaction rates. These are the results presented in Section \ref{sec:finalabuns}. A correlation analysis of the data is also used to identify key rates, as explained in Section \ref{sec:correlations}, with the results shown in Section \ref{sec:key}.

It should be kept in mind that the required computational time is largely independent of the number of varied reaction rates. Rather, it is determined by the time taken to follow the reaction network through a given trajectory, the number of trajectories, and the number of MC iterations required to obtain statistical significance. The solution time for the network is determined by the network size, i.e., the number $m$ of nuclei considered and the number $n$ of reactions connecting them. Each MC iteration initially sets up the vector $v_z$ for the variation and then runs the trajectory. That means that in each run all $z$ rates are varied. Choosing a sufficiently large number of iterations ensures that the space of possible combinations is largely sampled with a comparatively small number of variations. Using test calculations we confirmed that 10000 variations, i.e., MC iterations, are sufficient. This also means that each rate is varied 10000 times. With a given reaction network and trajectory, this requires always the same time, regardless of the size of $z$.

\subsection{Uncertainties}
\label{sec:uncert}

\subsubsection{Definitions}
\label{sec:errdefs}

It is not straightforward to define the uncertainties inherent in model calculations. This goes even beyond the distinction between statistical and systematic errors as used, e.g., in assigning error bars to measurements. This has been discussed, e.g., in \citet{sensi}. Two main sources of errors can be identified, the model itself and its input. Whenever input is derived from experimental data it will carry their statistical and systematic errors, which will be propagated through the model and into the final results. The error introduced by the choice of model, on the other hand, is difficult to quantify as the associated error is not statistical. Rather, the underlying assumptions have to be examined and an estimation has to be made about the inherent uncertainty. This estimate can be checked, to a certain extent, by comparison of the final results to further data but in principle it is impossible to ascertain the validity of a model completely \citep{Popper}. The case is complicated by the fact that some of the input may not be measured properties but will also come from a theoretical treatment within further models.

The usual approach for obtaining model uncertainties is to ignore possible errors in the main model and focus on the error propagation of input uncertainties. In this work, we follow a similar strategy, propagating uncertainties in the nuclear input, i.e., in the astrophysical reaction rates, through a given postprocessing model. Additionally, however, we compare the results from two different stellar evolution and explosion treatments. This also provides clues on the underlying model uncertainties.

Concerning uncertainties in stellar reaction rates, it is unavoidable to again combine experimental and theoretical uncertainties. This is partly due to the fact that cross sections of many of the reactions involved in the $\gamma$ process have not been measured, because either the target nuclei are radioactive or the Coulomb barrier prevents an experimental determination at the relevant, low energies. Even if a reaction cross section is measured, however, it still may have to be supplemented by predictions in order to arrive at a \textit{stellar} cross section $\sigma^*$ and thus a \textit{stellar} reactivity $\langle \sigma^* v \rangle$ as required in equations (\ref{eq:network})--(\ref{eq:revphoto}). Current laboratory measurements of reaction cross sections can only address target nuclei in their ground state and thus reactions proceeding on their excited states have to be treated in a reaction model. (This is not to say that experimental information on certain transitions leading to excited states or to test a reaction model cannot be extracted but theory has to be invoked to cast these data into an astrophysical reaction rate.)

Excited state contributions to the stellar rate depend on the plasma temperature and the spectroscopic properties of the involved nuclei. In general, a higher intrinsic nuclear level density in a nucleus will lead to a higher importance of excited state contributions but the presence of low-lying excited states, a few tens of keV above the ground state, may render them important even in nuclei with otherwise low level density. While for light nuclei with their large level spacings and high particle separation energies, excited state contributions are sizeable only for a limited number of cases, the inclusion of reactions on thermally populated target states has been shown to be important in all nucleosynthesis environments for the production of nuclei beyond Fe \citep{nosef,stellarerrors}.

The total contribution $X_\mathrm{exc}$ of reactions on thermally excited states of a nucleus can be quantified as
$X_\mathrm{exc}=1-X_0$, where $X_0$ is the ground state (g.s.) contribution \citep{nosef,stellarerrors}
\begin{equation}
\label{eq:xfactor}
X_0= \frac{2J_0+1}{G(T)} \frac{\langle \sigma^\mathrm{g.s.}_{Aa} v \rangle}{\langle \sigma^*_{Aa} v \rangle}\quad .
\end{equation}
Here, $\sigma^\mathrm{g.s.}$ is the reaction cross section for the target nucleus in the g.s.\ (this is the usual laboratory cross section) and $J_0$ is its g.s.\ spin.
It is very important to note that this definition of the g.s.\ contribution is different from the simple ratio of g.s.\ and stellar rates, respectively. Exhaustive tables of g.s.\ contributions are found in \citet{stellarerrors,sensi}.

An important application of g.s.\ contributions is the determination of rate uncertainties. Laboratory measurements of reaction cross sections only constrain $\sigma^\mathrm{g.s.}$ and the contributions of excited states to the stellar rate have to be determined from theory. Depending on $X_0$, the measured cross section will contribute more or less to the stellar rate. Likewise, the uncertainty in the stellar rate will be more or less dominated by the experimental error. \citet{stellarerrors} has shown that for a number of intermediate and heavy nuclei, $X_0$ is small already at $s$ process temperatures. In explosive conditions, such as for the $\gamma$ process with typical temperatures between 2 and 3.5 GK, $X_0$ will be small for most nuclei. Therefore it is necessary to construct an uncertainty that varies with temperature, according to the varying contributions of g.s.\ and excited states. Following \citet{stellarerrors,advances}, an experimental uncertainty factor $U_{\mathrm{exp}}$ for the reaction cross section of a target nucleus in the g.s.\ and a theoretical one $U_\mathrm{th}$  for the prediction of reactions with the target nucleus being excited can be combined to a total uncertainty $u^*$ of the stellar rate, with
\begin{equation}
u^*(T)=U_{\mathrm{exp}}+(U_\mathrm{th}-U_{\mathrm{exp}})X_\mathrm{exc}(T)=U_{\mathrm{exp}}+(U_\mathrm{th}-U_{\mathrm{exp}})(1-X_0(T)) \label{eq:uncertainty} \quad,
\end{equation}
assuming $U_\mathrm{th}>U_{\mathrm{exp}}$.
When there is no measurement, obviously the above equation trivially reduces to just the theory uncertainty.

As discussed in \citet{stellarerrors,sensi}, the predicted g.s.\ contributions themselves carry an uncertainty and it would be possible to include it in the above equation. As was also shown, however, it is small and does not affect the main impact, i.e., within their errors small contributions remain small, large ones remain large and the overall uncertainty $u^*$ is not largely changed. Therefore, we have ignored the uncertainty in $X_0$ when constructing our temperature-dependent uncertainties.

\subsubsection{Variation factors}
\label{sec:varfacts}

The PizBuin MC framework offers several ways of treating uncertainties. The possibility of using temperature-dependent errors has already been mentioned above. In addition, the uncertainties can be asymmetric around a given ``standard'' rate and the upper and lower envelopes can be set independently to any desired function of temperature. Random values can be drawn from three possible distributions: the uniform distribution, Gaussian distribution, and lognormal distribution. An individual uncertainty can be chosen for each rate although in practice it will often be the case that certain types of rates are assigned similar uncertainties in a region of nuclides.

When setting up the MC runs, uncertainties have to be chosen and the trajectories provided for the network calculation. During the actual runs, in each MC iteration a set of variation factors $v_z$ for the $z$ rates to be varied is randomly generated according to the preset variation type. A trajectory is processed with the factors applied to the rates (and applying the same factor to forward and reverse rate as explained in Section \ref{sec:errdefs}) and the resulting abundances are stored. Then a new iteration is started with a newly chosen set of variation factors.

In a Gaussian distribution, the function values $\mathcal{Y}$ are normally distributed, while in a lognormal distribution $\log \left(\mathcal{Y}\right)$ are normally distributed.
While historically the Gaussian distribution has been mostly used in statistical error analysis, recently the lognormal distribution is viewed as being a more appropriate choice for natural processes, especially when many unknown factors are acting together \citep{jaynes,mc1}. Where Gaussian distributions are added and therefore account for addition
of contributing terms, lognormal distributions are multiplied and
stem from multiplication of contributing factors. Nowadays, the lognormal
distribution is used in many different fields, from biology to finance \citep{mc1}.
For experimental errors and variation of rates
and cross sections, it has the beneficial feature that it implicitly forbids the results to assume
unphysical, negative values. We will see in Section \ref{sec:finalabuns} that abundance uncertainties assume distributions very close to lognormal distributions even when the uncertainties in the input were not lognormally distributed. This arises from the error multiplication effect when several reactions are contributing to the total uncertainty.

Gaussian and lognormal distributions are related but a determination of the usually used (Gaussian) confidence intervals is  complicated and non-trivial for the general case of a lognormal distribution.
One approach is to extract confidence limits from the underlying normal
distribution and convert them to (asymmetric) ones of the lognormal
distribution. A straightforward approach only works if the derived
confidence interval does not include zero or negative values. For
a sample recipe, see, e.g., \citet{SmiNaber}.

In astrophysics, sensitivity studies usually vary rates by a certain factor smaller or larger than unity, to check how the abundance of a nuclide produced in a nucleosynthesis process depends on a rate. This would also be an appropriate way to test dependence on theoretical models, which cannot be assumed to be drawn from a statistical distribution but rather exhibit only systematic uncertainties \citep{sensi}. Such a variation can be modeled by using variation factors drawn from a uniform distribution, which generates values with equal probability between two limits. Since in this first application of our MC code we want to combine experimental and theoretical errors, we also assumed such uniform distributions for experimental errors. Gaussian errors on measured values were converted by taking 2$\sigma$ confidence intervals to define the boundaries of the value range for the uniform distribution.

The theoretical uncertainties should include realistic assumptions for the uncertainties introduced by the input for a stellar rate calculation, as well as for the systematic uncertainty by the application of specific models to treat input and cross section calculations. These were motivated by comparisons between experimental and predicted g.s.\ cross sections across the nuclear chart and are shown in Table \ref{tab:defineuncert}.

For example, \citet{rtk97} found an average deviation of 30\% for 30 keV neutron captures along stability but also show systematic deviations of up to a factor of two in certain mass regions, especially close to magic numbers where the Hauser-Feshbach model reaches its limit. Although this limitation is at least partially lifted at higher plasma temperatures, we nevertheless assumed a factor of two uncertainty in predicted (n,$\gamma$) rates.

On the other hand, recent experiments have shown that the standard rates by \cite{adndt00} systematically overestimate ($\alpha$,$\gamma$) rates \citep[see, e.g.,][and references therein]{p-review}. The deviations found are between factors 1.0 and 0.1 . Accordingly, we use an asymmetric error for such rates, allowing only limited variation to higher values but setting the lower bound to 0.1 of the standard rate.

\begin{table}
        \centering
        \caption{Theory uncertainty factors assumed for various reaction types; for the upper limit, the rate is multiplied by $U_\mathrm{th}^\mathrm{hi}$, for the lower limit it is divided by $U_\mathrm{th}^\mathrm{lo}$. For measured g.s.\ cross sections, these uncertainties were combined with the experimental error to derive a temperature-dependent uncertainty factor as shown in equation (\ref{eq:uncertainty}). Note that forward and reverse rates, e.g., captures and photodisintegrations, are modified by the same factor.}
        \label{tab:defineuncert}
        \begin{tabular}{lrr} 
                \hline
                Reaction & $U_\mathrm{th}^\mathrm{hi}$ & $U_\mathrm{th}^\mathrm{lo}$ \\
                \hline
                (n,$\gamma$) & 2.0 & 2.0 \\
                (p,$\gamma$) & 2.0 & 3.0 \\
                (p,n) & 2.0 & 3.0 \\
                ($\alpha$,$\gamma$) & 2.0 & 10.0 \\
                ($\alpha$,n) & 2.0 & 10.0 \\
                ($\alpha$,p) & 2.0 & 10.0 \\
                \hline
        \end{tabular}
\end{table}

Recent investigations have found that the proton widths at low energy predicted using the optical potential of \citet{jlm} in the version of \citet{lejeune}, as in the standard theoretical rate set used here, may be slightly overpredicted. A modification of the optical potential leads to a better reproduction of experimental data \citep{Kiss08,Rauscher09,Netterdon16}. To account for this, we introduce an asymmetric uncertainty also for reactions involving protons in the entrance or exit channel and allow for a larger variation towards lower values of the reaction rate.

We included rates on target nuclei from Fe to Bi in our variations, including all unstable nuclei involved in the $\gamma$ process. The full network thus included about 3800 nuclei, each involved in several reaction types. Although the assigned uncertainties were motivated by comparisons for stable nuclei, the $\gamma$-process does not involve nuclei far from stability and thus it is a reasonable assumption that the same theory uncertainties also apply, especially because we already chose conservative limits on those.

Uncertainties for reactions involving the weak interaction -- $\beta^-$, $\beta^+$, and electron captures -- also included a temperature dependence but this was treated slightly differently than described in equation (\ref{eq:uncertainty}) because $X_0$ is not available for these reactions. The theory uncertainty for predicting weak reactions on excited states was assigned a value $^\beta U_\mathrm{th}= {^\beta U_\mathrm{th}^\mathrm{hi}}= {^\beta U_\mathrm{th}^\mathrm{lo}}=10$ and the temperature-dependent uncertainty $u_\beta ^*$ was computed with
\begin{equation}
u_\beta ^* = \frac{\left( 2J_0+1\right) {^\beta U_{\mathrm{g.s.}}}}{G(T)}+ {^\beta U_\mathrm{th}}\left( 1 - \frac{2J_0+1}{G(T)}  \right) \quad, \label{eq:betauncert}
\end{equation}
where $^\beta U_{\mathrm{g.s.}}$ is either the uncertainty of a measured half-life or a factor 1.3. Thus, these rates become more uncertain with increasing temperature. Uncertainties in half-lives do not play a crucial role in the $\gamma$ process. They are important, however, in the main and weak $s$ process and are further discussed, e.g., in \citet{weakNobuya}.

Having chosen the uncertainty range $[r^*_0(T)/u^*_\mathrm{lo}(T),r^*_0(T) u^*_\mathrm{hi}(T)]$ as described above and the probability distribution for each reaction to be varied, the random value $v_{z\rightarrow j}$ supplied by the MC driver is mapped to an actual variation factor $f_j(T)$ of the $j$-th rate according to
\begin{equation}
f_j(T)=\frac{1}{u^*_\mathrm{lo}} + v_{z\rightarrow j} \left( u^*_{\mathrm{hi}}-\frac{1}{u^*_{\mathrm{lo}}} \right) \quad . \label{eq:varfact}
\end{equation}
This means that the rate value $r^*_\mathrm{new}(T)$ used in the reaction network is given by $r^*_\mathrm{new}(T)=f_j(T) r^*_0(T)$, where $r^*_0(T)$ is the original, unchanged stellar rate as given in the reaction rate library.

\subsection{Correlations to identify key rates}
\label{sec:correlations}

In past variation studies key reactions were identified manually by inspection of nuclear flows and by comparing the impact of the variation of a few reactions on final abundances. This is not possible with complicated flow patterns and many contributing rates. Moreover, the impact of reactions can be modified when accounting for the uncertainties of all involved reactions because variations in apparent key reactions can be partially or fully suppressed by combined variations of other rates. Therefore, limiting the investigation to a few reactions may be misleading.

\begin{figure}
\includegraphics[width=\columnwidth]{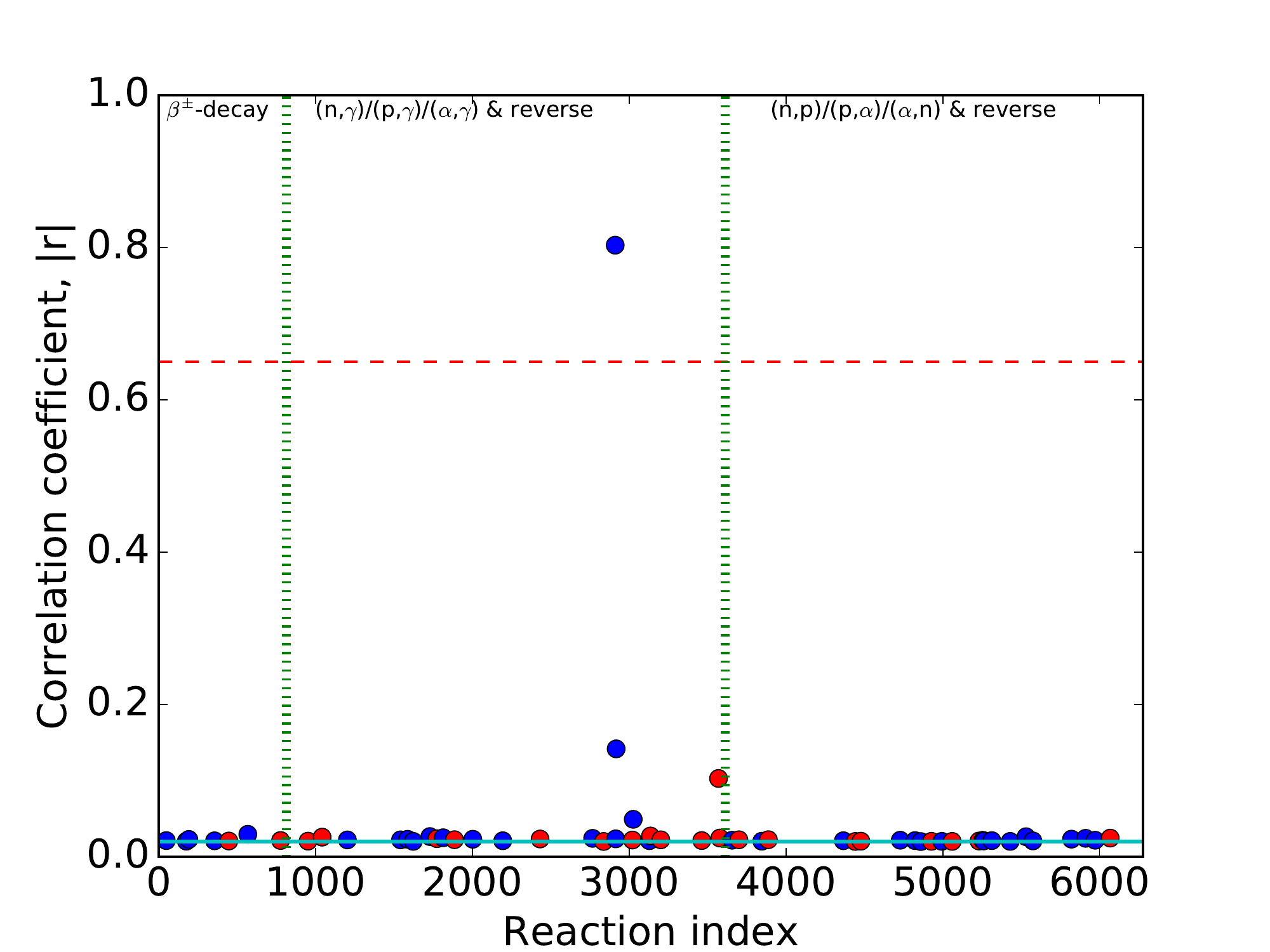}
\caption{Example for the weighted correlation values obtained after the initial MC run for reactions with respect to a specific nuclide.
All rates are numbered and shown on the horizontal axis. Correlations above 0.65 are considered to be strong. Red dots imply $r>0$, blue dots are $r<0$ (inverse correlation). Correlation values below 0.02 (cyan line) are not shown. Typically, only one rate is strongly correlated, if any at all.
A strong correlation after the initial MC run defines level 1 key rates.\label{fig:corr1}}
\end{figure}


The stored MC data allow for a more comprehensive approach and a fully automated search for actual key rates. Since the variation factors for each rate are found in the stored variation vectors $v_z$, it can be tested whether there is a correlation between the variation of a rate and the resulting change in abundance. The correlation will be larger the fewer reactions contribute to the uncertainty of a given abundance. There are various definitions for correlations in the literature. We employ a widely-used correlation coefficient, the Pearson product-moment correlation coefficient \citep{pearson}
\begin{equation}
\label{eq:pearson}
r_\mathrm{Pearson}=\frac{\sum_{i=1}^k\left( \tilde{x}_i -\overline{x} \right)\left( \tilde{y}_i - \overline{y}\right) }{\sqrt{\sum_{i=1}^k\left( \tilde{x}_i -\overline{x} \right)^2}\sqrt{\sum_{i=1}^k\left( \tilde{y}_i -\overline{y} \right)^2}}
\end{equation}
for samples of two datasets $\tilde{x}_i,\tilde{y}_i$ of $k$ values, with the sample means of $\overline{x}=\left(\sum_{i=1}^k \tilde{x}_i \right)/k$, $\overline{y}=\left(\sum_{i=1}^k \tilde{y}_i \right)/k$. In our case, $\tilde{x}_i$ are variation factors and $\tilde{y}_i$ abundances. The number of variation factors, i.e., the number of MC iterations is denoted by $k$.

Since we are interested in key rates which globally affect the final abundances and not just those in one trajectory, it was necessary to modify equation (\ref{eq:pearson}) to provide a weighted average over all trajectories used. Thus, our weighted correlation $r_\mathrm{corr}$ is given by
\begin{equation}
\label{eq:weightcorr}
r^q_\mathrm{corr}=\frac{\sum_{ij} {^qw_{j}^2} \left( f_{ij} -\overline{f}_j \right)\left( ^qY_{ij} - \overline{^qY}_j\right) }{\sqrt{\sum_{ij} {^qw_j^2}\left( f_{ij} -\overline{f}_j \right)^2}\sqrt{\sum_{ij} {^qw_j^2} \left( ^qY_{ij} -\overline{^qY}_j \right)^2}} \quad.
\end{equation}
Here, the trajectory is given by $j$ and the iteration by $i$, with variation factor $f_{ij}$ of the rate and final abundance $^qY_{ij}$ of nuclide $q$ resulting from this variation. To connect all rates to all abundances of interest, $z$ weighted correlation factors have to be computed for each nuclide of interest. The weight of each trajectory is calculated from the relative abundance change
\begin{equation}
\label{eq:normprod}
^qw_j=\frac{|^qY^\mathrm{std}_j-^qY^\mathrm{ini}_j|}{\sum_j {|^qY^\mathrm{std}_j-^qY^\mathrm{ini}_j}|}
\end{equation}
for each nuclide $q$ with initial abundance $^qY^\mathrm{ini}_j$ in trajectory $j$. The final abundances obtained with the standard rate set are denoted by $^qY^\mathrm{std}_j$. The summations in equation (\ref{eq:weightcorr}) can be made more economically when realizing that for most $p$ nuclei, only very few trajectories contribute and thus have non-negligible weights (see Section \ref{sec:stellarmodels}).

The weighted correlations as defined above result in values $0\leq \left|r^q_\mathrm{corr}\right| \leq 1$, with 0 meaning no correlation. Negative values show an anti-correlation, i.e., an increase in the rate would lead to a decrease in abundance and vice versa. In order to understand and interpret correctly the weighted correlations, it is useful to keep several things in mind. Very importantly, the weighted correlation in itself is not an absolute measure of the impact of a rate variation as its value depends on the number of varied rates. For example, when only a single rate is varied, the correlation will always be 1, even if this rate does not change the abundance. It immediately follows that values of the weighted correlations can only be compared within a set of MC iterations without change in the number of varied rates. A cross-comparison between different MC sets with different numbers of varied rates is meaningless.

The above considerations also imply that the obtained correlation values cannot be used directly to rank the important rates. As an example, consider the situation where the uncertainty of the rate with the highest correlation value has been removed by a measurement, it would be taken out of the MC variation and different correlation values will appear for the remaining rates. Not just the absolute values, however, will be modified but also the relative values of the rates compared to each other can be changed, i.e., their relative ranking. This is because the large uncertainty in a key rate can mask the impact of another rate. When the key rate is taken out of the variation (which implies that it has zero uncertainty), it will not be able to compensate another rate uncertainty and another rate may appear with a large uncertainty correlation. Examples for this are seen in the final results presented in Section \ref{sec:key}. It is an advantage of this MC correlation method over the manual variation of a few, individual rates because the latter cannot capture the mutual compensations of uncertainties in the simultaneous variation of many rates.

Review of the available literature suggests that a Pearson product-moment correlation coefficient value above 0.7 indicates a strong correlation. We choose a threshold of 0.65 to account for numerical uncertainties in our calculations. In our identification of key rates for the production of each $p$ nucleus, we perform several steps. If rates with a value above 0.65 are present in our initial MC run, we define these as key rates of high importance. Then another MC run is performed without variation of those key rates. If rates with correlations above 0.65 appear again, we take these as rates of second level importance, i.e., they will be of interest as soon as the previously found key rates have been determined with higher accuracy. Another such step is taken to identify third level key rates. If no key rates are found in any of these steps, no further step is taken. This means that the remaining uncertainties in the final abundances are determined by the combined uncertainties of several rates.

Figure \ref{fig:corr1} shows an example of the weighted correlations found for a nuclide. Typically, only a few reactions show sizeable correlation, if any at all, while the majority have negligible correlation.

It is important to remember that this definition of key rate selects the rates that are mainly contributing to the uncertainty in the production of a given nucleus. In other words, the uncertainty of a key rate cannot be compensated for by variations in other rates. This does not necessarily imply that an abundance is very sensitive to the rate, i.e., that a small rate variation results in a large abundance change. For instance, it is conceivable that an abundance being sensitive to a rate with small uncertainty can be affected more by another rate with larger uncertainty to which it is less sensitive.

\begin{figure}
\includegraphics[width=\columnwidth]{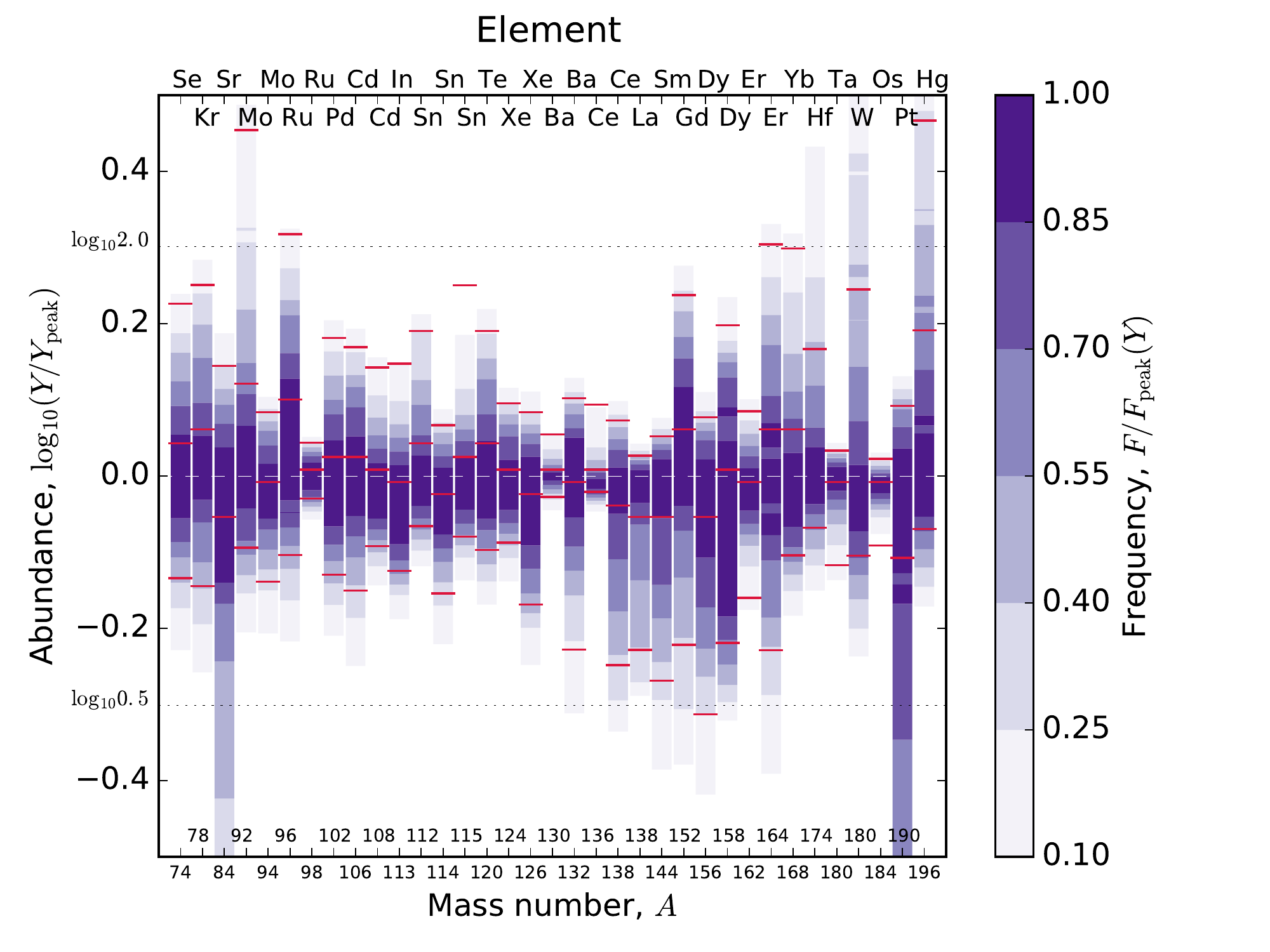}
\caption{Total production uncertainties of the classical $p$ nuclides in the explosion of a 25 $M_\odot$ solar metallicity star, obtained with trajectories from \citet{hashi}. The color shade is the probabilistic frequency and the 90\% probability intervals up and down marked for each nuclide (see Fig.\ \ref{fig:exampledistrib} for further details). Horizontal dashed lines indicate a factor of two uncertainties.}
\label{fig:hashi25}
\end{figure}

\begin{figure}
\includegraphics[width=\columnwidth]{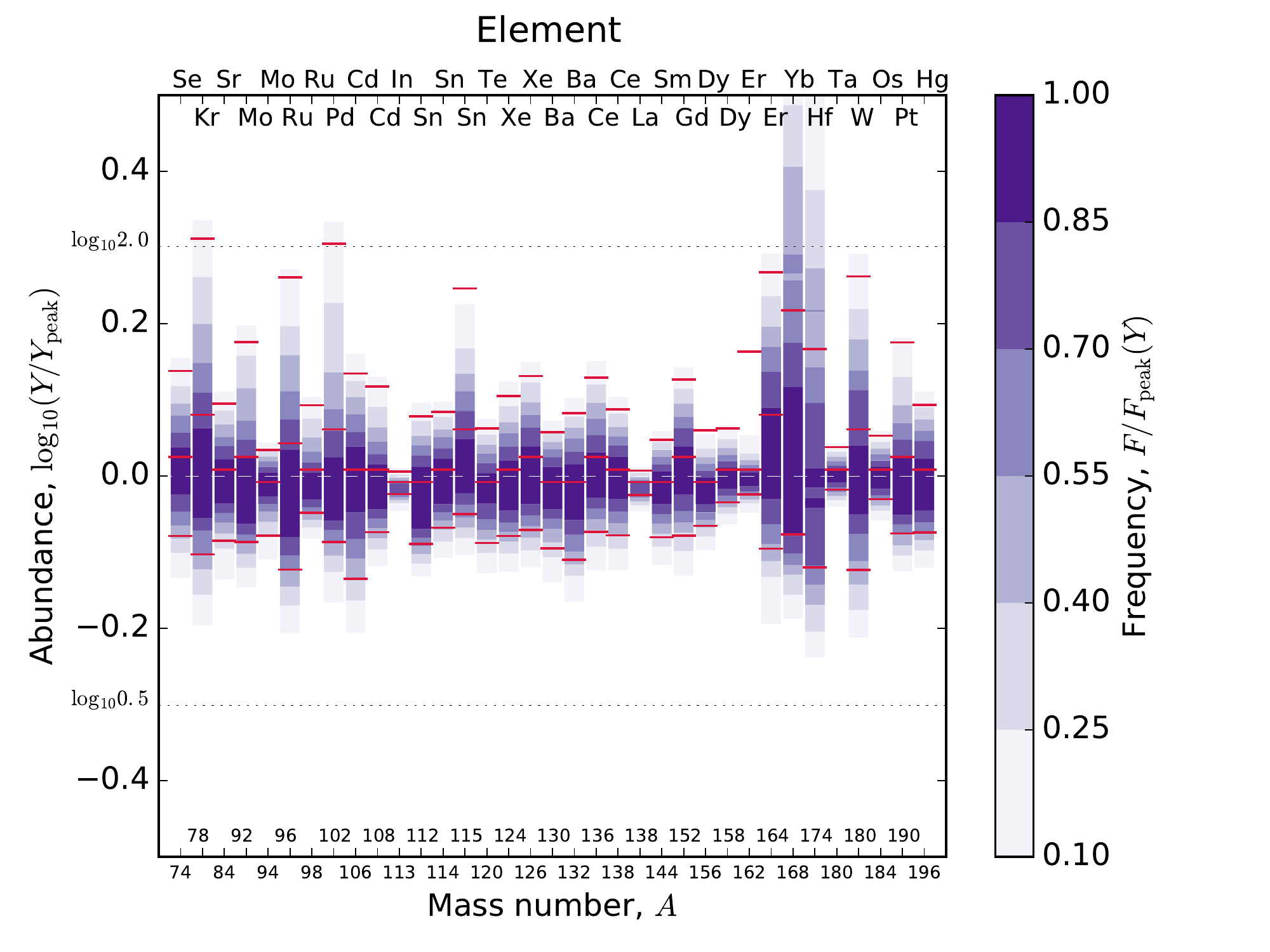}
\caption{Same as Fig.\ \ref{fig:hashi25} but for the explosion of a 25 $M_\odot$ solar metallicity star, obtained with KEPLER trajectories. \label{fig:kepler25}}
\end{figure}

\begin{figure}
\includegraphics[width=\columnwidth]{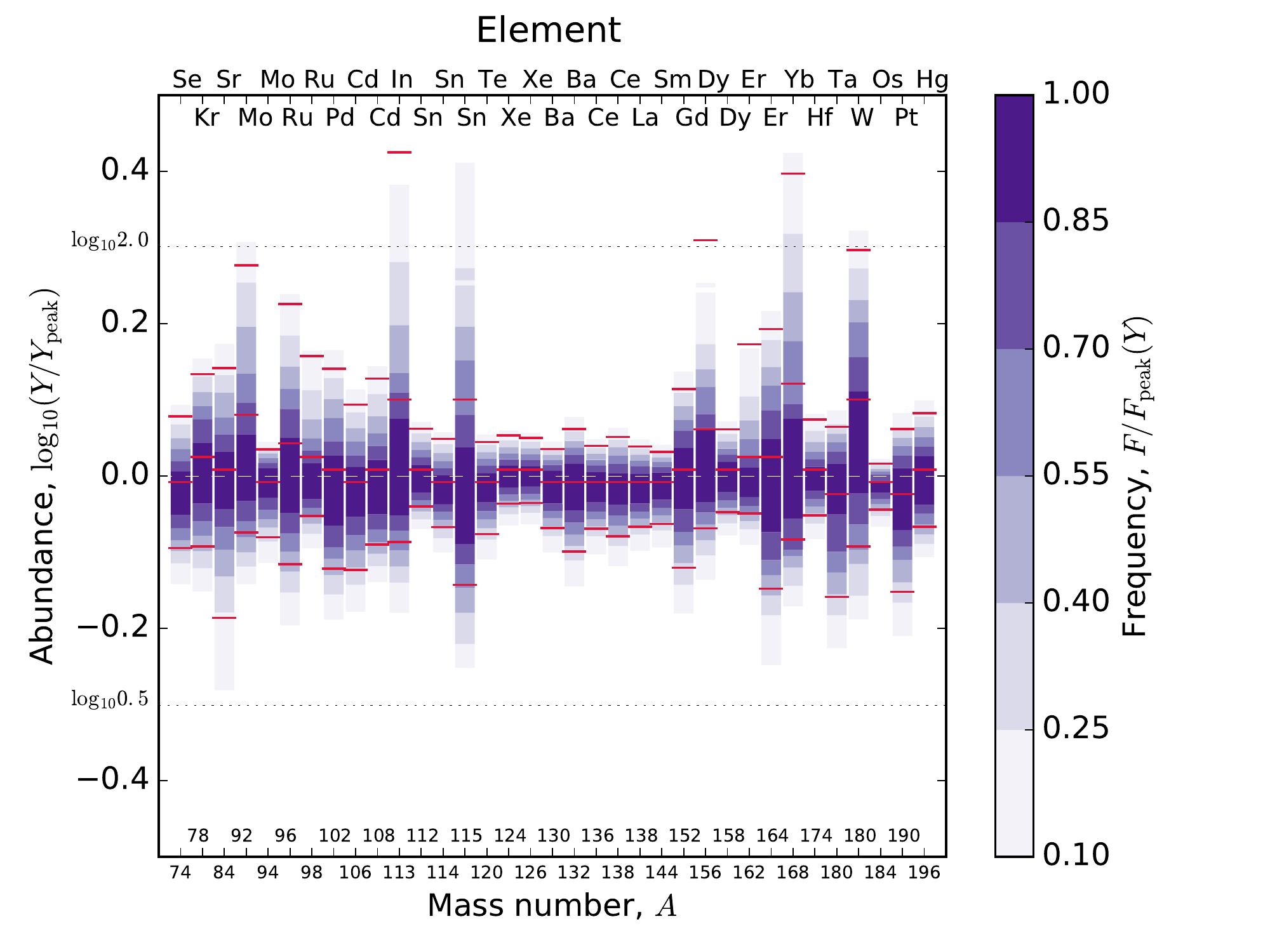}
\caption{Same as Fig.\ \ref{fig:kepler25} but for the explosion of a 15 $M_\odot$ solar metallicity star. \label{fig:kepler15}}
\end{figure}

\begin{figure}
\includegraphics[width=\columnwidth]{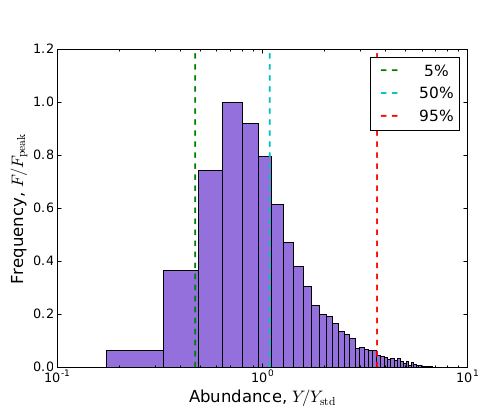}
\caption{Example of the abundance distributions obtained in the MC runs. The bounds encompassing 5\% and 95\% of the distribution are marked to allow to use them as uncertainty measures. \label{fig:exampledistrib}}
\end{figure}

\section{Results}

The standard rate set used was based on the theoretical rates by \citet{adndt00}, supplemented by experimental rates taken from \citet{kadonis,cyburt}. Decays and electron captures were taken from a REACLIB file compiled by \citet{freiburghaus} and supplemented by rates from \cite{taka,gor99} as provided by \citet{NetGen05,NetGen13}. These rates were varied according to the temperature-dependent uncertainties assigned as described in Section \ref{sec:varfacts}.

Since the $p$-nucleus production depends on the mass of the supernova progenitor \citep{woohow,rhhw02}, we chose trajectories and initial abundances obtained in a 15 $M_\odot$ and a 25 $M_\odot$ model evolved in the KEPLER stellar evolution code \citep{kepler,rhhw02,hegwoo07} starting from abundances by \cite{lodders,lodders2}. Trajectories and initial abundances from the O/Ne shell of a 25 $M_\odot$ model by \citet{hashi} have been frequently used in earlier post-processing studies with manual variation of reaction rates for the $\gamma$ process \citep[see, e.g.,][]{rayet95,rapp,rauscherbranch,pign-rev}. Therefore, we also studied final abundance uncertainties obtained in that model for comparison.

\begin{table}
        \centering
        \caption{Total production uncertainties from MC postprocessing of the 25 $M_\odot$ model of \citet{hashi}. The shown uncertainty factors for variations up and down enclose a 90\% probability interval, as shown in Fig.\ \ref{fig:exampledistrib}.}
        \label{tab:hashiuncert}
        \begin{tabular}{lrr}
                \hline
                Nuclide  &  up &  down \\
                \hline
$^{74}{\rm Se}$&1.683&0.734 \\
$^{78}{\rm Kr}$&1.781&0.717 \\
$^{84}{\rm Sr}$&1.395&0.266 \\
$^{92}{\rm Mo}$&2.845&0.805 \\
$^{94}{\rm Mo}$&1.213&0.726 \\
$^{96}{\rm Ru}$&2.078&0.787 \\
$^{98}{\rm Ru}$&1.106&0.935 \\
$^{102}{\rm Pd}$&1.518&0.742 \\
$^{106}{\rm Cd}$&1.476&0.707 \\
$^{108}{\rm Cd}$&1.389&0.809 \\
$^{113}{\rm In}$&1.404&0.751 \\
$^{112}{\rm Sn}$&1.551&0.860 \\
$^{114}{\rm Sn}$&1.166&0.701 \\
$^{115}{\rm Sn}$&1.779&0.832 \\
$^{120}{\rm Te}$&1.550&0.799 \\
$^{124}{\rm Xe}$&1.246&0.818 \\
$^{126}{\rm Xe}$&1.212&0.678 \\
$^{130}{\rm Ba}$&1.134&0.938 \\
$^{132}{\rm Ba}$&1.265&0.591 \\
$^{136}{\rm Ce}$&1.241&0.953 \\
$^{138}{\rm Ce}$&1.183&0.565 \\
$^{138}{\rm La}$&1.064&0.591 \\
$^{144}{\rm Sm}$&1.127&0.539 \\
$^{152}{\rm Gd}$&1.728&0.600 \\
$^{156}{\rm Dy}$&1.194&0.487 \\
$^{158}{\rm Dy}$&1.577&0.603 \\
$^{162}{\rm Er}$&1.216&0.692 \\
$^{164}{\rm Er}$&2.014&0.590 \\
$^{168}{\rm Yb}$&1.990&0.787 \\
$^{174}{\rm Hf}$&4.875&0.855 \\
$^{180}{\rm Ta}$&1.080&0.763 \\
$^{180}{\rm W}$&5.156&0.786 \\
$^{184}{\rm Os}$&1.054&0.811 \\
$^{190}{\rm Pt}$&1.237&0.221 \\
$^{196}{\rm Hg}$&2.929&0.852 \\
\hline
        \end{tabular}
\end{table}

\begin{table}
        \centering
        \caption{Same as Table \ref{tab:hashiuncert} but for the 15 and 25 $M_\odot$ KEPLER models.}
        \label{tab:kepleruncert}
        \begin{tabular}{lrrrr}
                \hline
                &\multicolumn{2}{c}{15 $M_\odot$ model} & \multicolumn{2}{c}{25 $M_\odot$ model} \\
                Nuclide &   up &  down &   up &  down\\
                \hline
$^{74}{\rm Se}$&1.198&0.804     &1.373&0.834       \\
$^{78}{\rm Kr}$&1.361&0.808     &2.049&0.789       \\
$^{84}{\rm Sr}$&1.386&0.652     &1.245&0.822       \\
$^{92}{\rm Mo}$&1.891&0.843     &1.499&0.819       \\
$^{94}{\rm Mo}$&1.084&0.831     &1.082&0.835       \\
$^{96}{\rm Ru}$&1.682&0.766     &1.823&0.753       \\
$^{98}{\rm Ru}$&1.437&0.886     &1.239&0.895       \\
$^{102}{\rm Pd}$&1.384&0.756    &2.018&0.819      \\
$^{106}{\rm Cd}$&1.240&0.753    &1.364&0.733      \\
$^{108}{\rm Cd}$&1.342&0.813    &1.310&0.844      \\
$^{113}{\rm In}$&2.660&0.819    &1.013&0.947      \\
$^{112}{\rm Sn}$&1.154&0.912    &1.198&0.814      \\
$^{114}{\rm Sn}$&1.119&0.857    &1.214&0.856      \\
$^{115}{\rm Sn}$&3.500&0.720    &1.764&0.891      \\
$^{120}{\rm Te}$&1.108&0.839    &1.155&0.816      \\
$^{124}{\rm Xe}$&1.130&0.920    &1.273&0.834      \\
$^{126}{\rm Xe}$&1.122&0.922    &1.353&0.850      \\
$^{130}{\rm Ba}$&1.085&0.854    &1.142&0.804      \\
$^{132}{\rm Ba}$&1.152&0.796    &1.209&0.776      \\
$^{136}{\rm Ce}$&1.095&0.853    &1.347&0.845      \\
$^{138}{\rm Ce}$&1.126&0.833    &1.223&0.836      \\
$^{138}{\rm La}$&1.093&0.858    &1.017&0.944      \\
$^{144}{\rm Sm}$&1.076&0.865    &1.116&0.831      \\
$^{152}{\rm Gd}$&1.300&0.758    &1.339&0.835      \\
$^{156}{\rm Dy}$&2.040&0.853    &1.148&0.860      \\
$^{158}{\rm Dy}$&1.152&0.897    &1.155&0.924      \\
$^{162}{\rm Er}$&1.490&0.893    &1.456&0.946      \\
$^{164}{\rm Er}$&1.559&0.712    &1.851&0.803      \\
$^{168}{\rm Yb}$&2.495&0.825    &3.476&0.838      \\
$^{174}{\rm Hf}$&1.186&0.888    &3.328&0.759      \\
$^{180}{\rm Ta}$&1.160&0.694    &1.091&0.960      \\
$^{180}{\rm W}$&1.981&0.808     &1.829&0.753      \\
$^{184}{\rm Os}$&1.038&0.903    &1.130&0.932      \\
$^{190}{\rm Pt}$&1.153&0.704    &1.498&0.841      \\
$^{196}{\rm Hg}$&1.209&0.858    &1.239&0.843      \\
\hline
\end{tabular}
\end{table}

Figures \ref{fig:hashi25} and \ref{fig:kepler25} show the uncertainties in production (or destruction) of the classical 35 $p$ nuclei obtained by our MC method in the 25 $M_\odot$ models by \citet{hashi} and KEPLER, respectively. These can be compared directly. The results for a 15 $M_\odot$ KEPLER model are shown in Fig.\ \ref{fig:kepler15}. The numerical values are given in Tables \ref{tab:hashiuncert} and \ref{tab:kepleruncert}, respectively.\footnote{The nuclei $^{138}$La and $^{180}$Ta receive strong contributions from the $\nu$-process which is not included in our postprocessing; the shown uncertainties only refer to other reactions involving these isotopes.}
The color shading in the figures provides the probabilistic frequency which counts how often an abundance value was obtained in the MC variations. 
Note that the uncertainty distributions are asymmetric, especially for the cases with large uncertainties. For easier legibility, the uncertainty bounds enclosing 90\% of the distribution are marked. Figure \ref{fig:exampledistrib} shows an example of such a distribution. As mentioned in Section \ref{sec:varfacts}, the more reactions contribute the closer it becomes to a pure lognormal distribution.

\subsection{Differences between stellar models}
\label{sec:stellarmodels}

\begin{figure}
\includegraphics[width=\columnwidth]{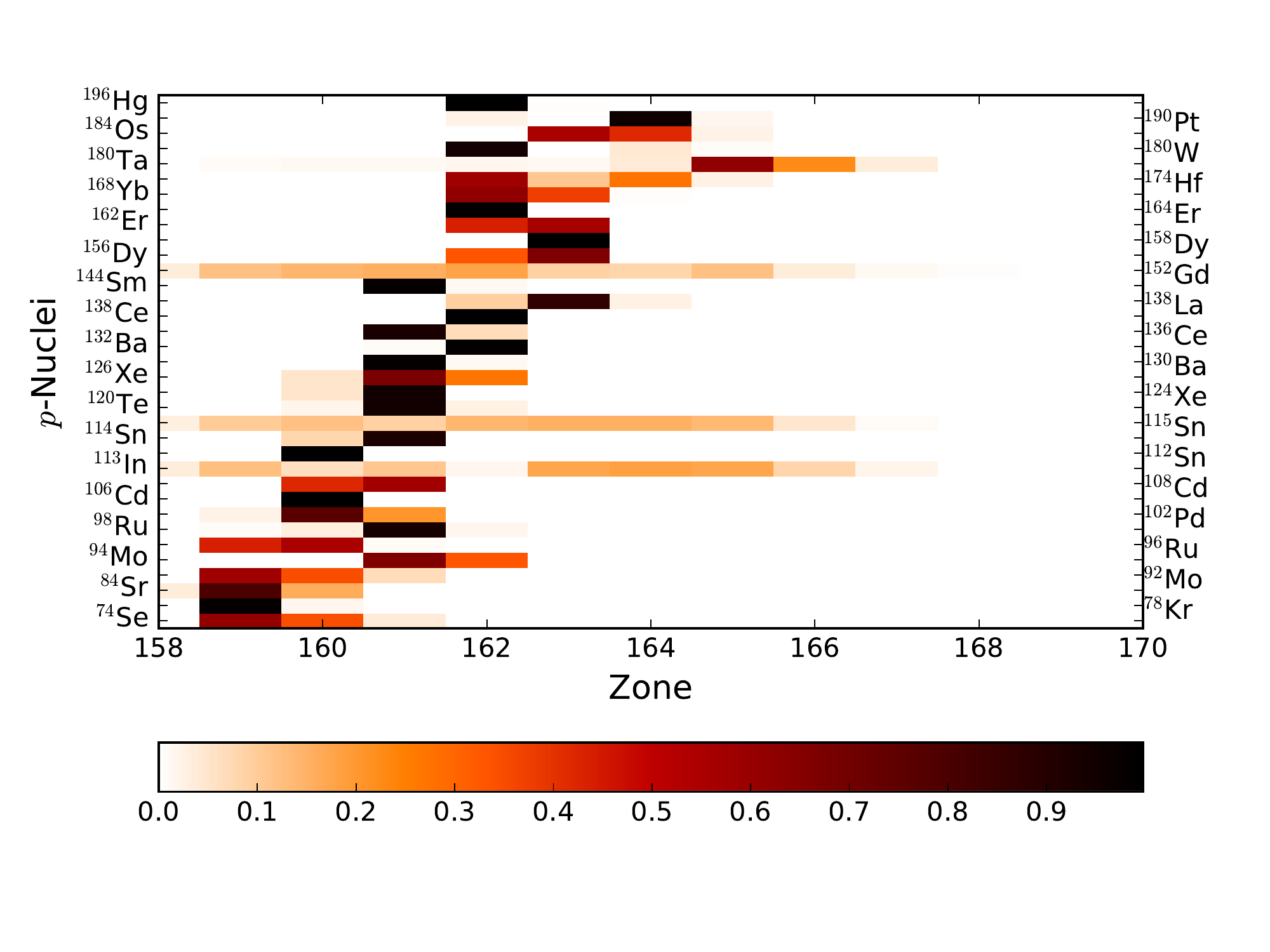}
\caption{\label{fig:peakhashi} Relative abundance change (as defined in equation \ref{eq:normprod}) for each nucleus of interest in each zone of the 25 $M_\odot$ model by \citet{hashi}. The (arbitrary) zone number is given on the abscissa; lower zone numbers are deeper inside the star.}
\end{figure}

\begin{figure}
\includegraphics[width=\columnwidth]{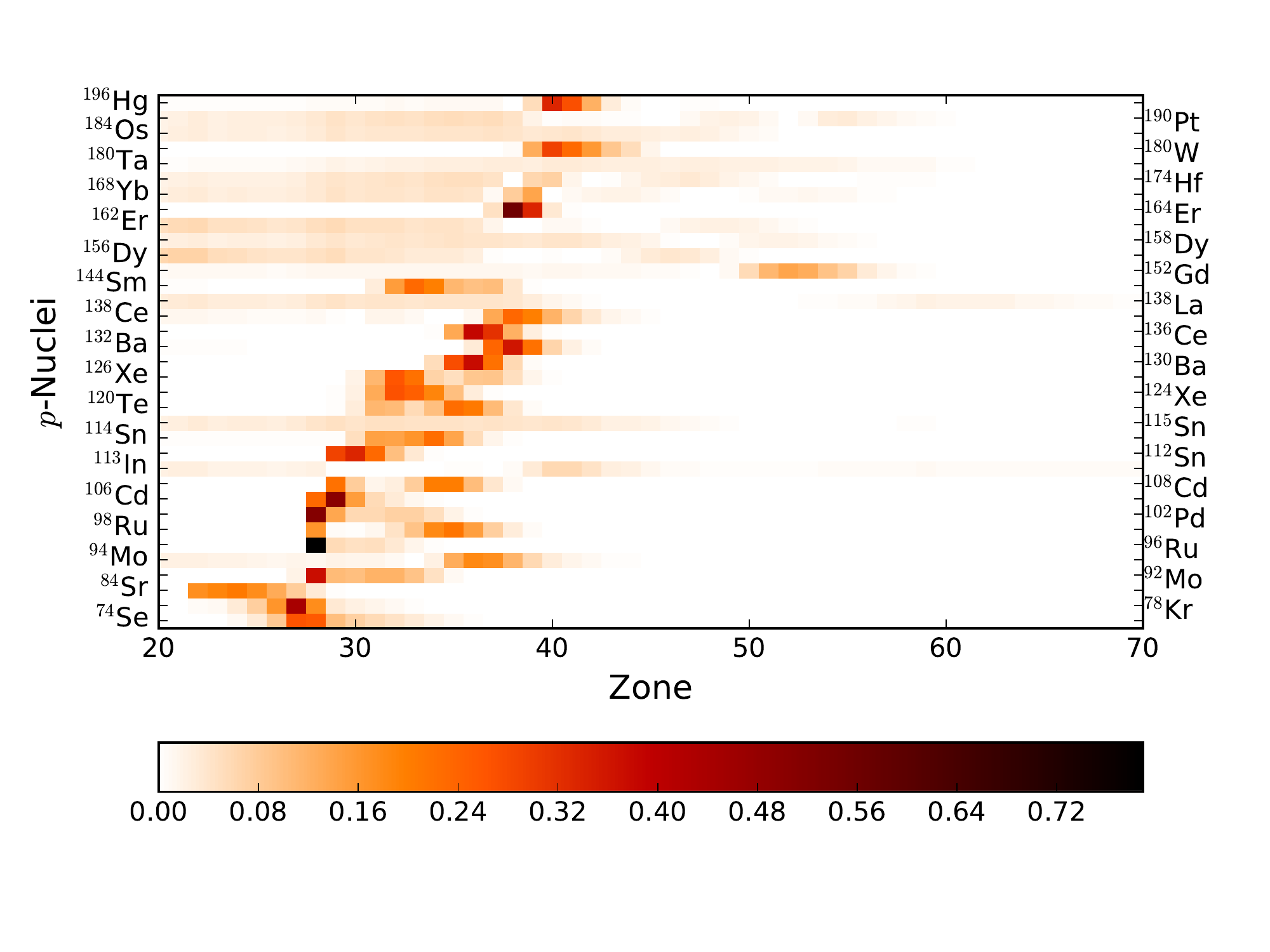}
\caption{\label{fig:peakK25} Same as Fig.\ \ref{fig:peakhashi} but for the 25 $M_\odot$ KEPLER model.}
\end{figure}

\begin{figure}
\includegraphics[width=\columnwidth]{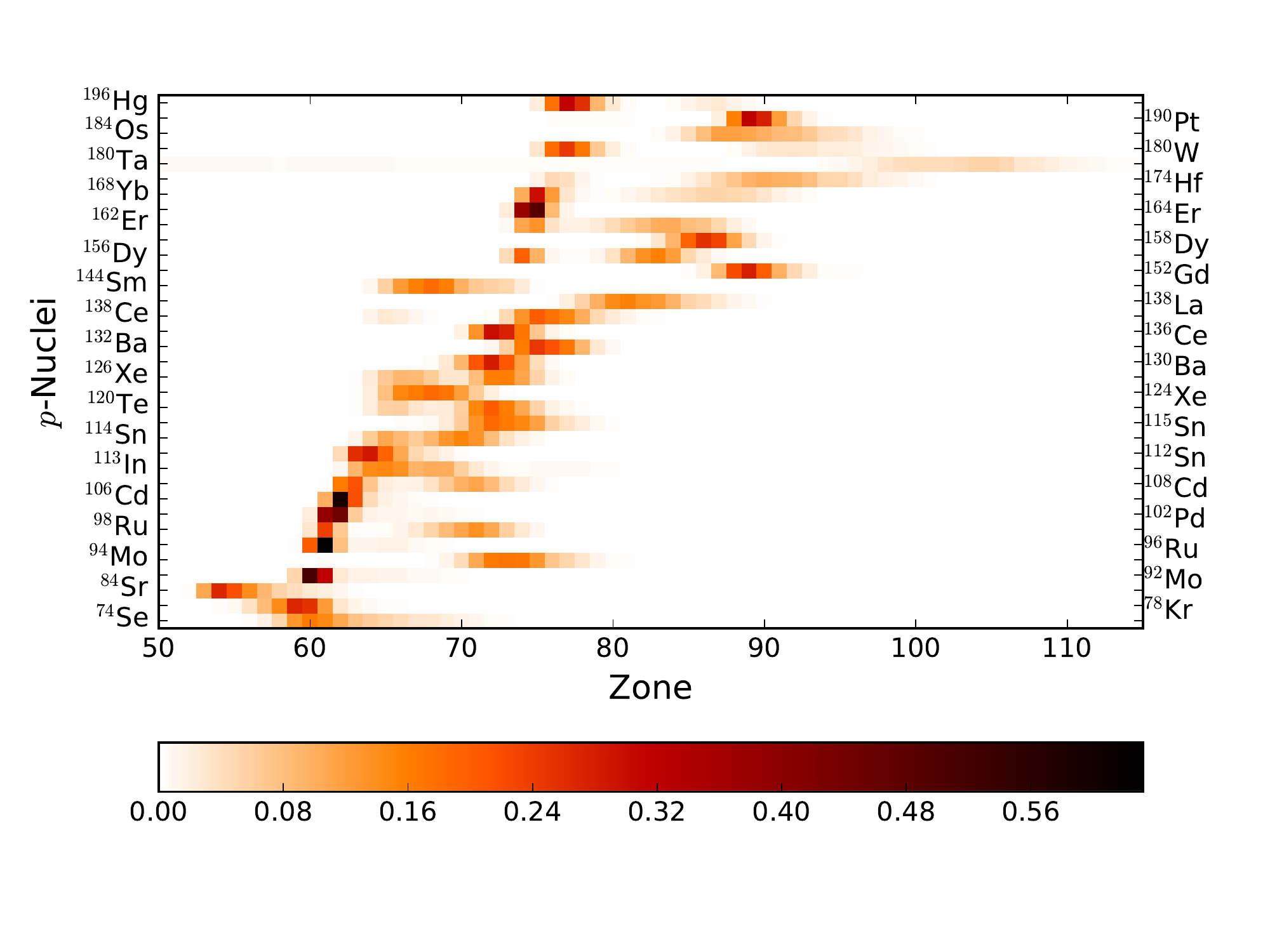}
\caption{\label{fig:peakK15} Same as Fig.\ \ref{fig:peakhashi} but for the 15 $M_\odot$ KEPLER model.}
\end{figure}

Although the main focus here is on nuclear uncertainties in $\gamma$-process calculations, it is worthwhile to also explore differences arising from the use of different codes, even when they attempt to model the same physical environment. As is evident from Figs.\ \ref{fig:hashi25} and \ref{fig:kepler25}, although the general pattern seems to be comparable, the overall uncertainty is found to be larger when using the \citet{hashi} trajectories than the ones obtained with KEPLER trajectories. For several nuclei, however, there are distinct differences between the uncertainties obtained in each calculation. The differences in uncertainty magnitude and for specific isotopes are not so much due to the possibility of differences in calculated stellar evolution up to the point when the supernova shock is passing through the O/Ne layer of the star but, rather, can be traced back to differences in the applied grid of mass zones. For both models, each trajectory is defined by the hydrodynamical evolution of a single mass zone of the stellar model. Each zone experiences a rapid increase in temperature and density when the supernova shock wave moves through. The shockwave loses energy on its way outwards and thus the peak temperature reached will be lower in zones located further out. By definition, all material in a zone is exposed to the same temperature and density history. A finer grid of mass zones thus will sample a finer graduation in peak temperatures. The achieved peak temperature is an essential factor in the production of $p$ nuclei because they are produced only in a narrow range of temperatures. Trajectories with too high $T$ will destroy too much, while those with too low $T$ will not be able to change abundances significantly. Furthermore, different mass ranges of $p$ nuclei require different temperatures. Abundance changes for the light $p$ nuclei are only possible at higher $T$ (around 3 GK) because their nuclear binding is larger.

Figures \ref{fig:peakhashi} -- \ref{fig:peakK15} show the relative production (or destruction) of nuclei in each mass zone. The zoning grid in the model of \citet{hashi} (Fig.\ \ref{fig:peakhashi}) is coarser than the one in \citet{rhhw02} (Figs.\ \ref{fig:peakK25}, \ref{fig:peakK15}), each zone encloses more mass and extends over a larger spatial distance. A coarser grid samples a coarser distribution of peak temperatures since more stellar material experiences a similar evolution of temperature and density. With respect to nucleosynthesis, not all relevant reactions and reaction flows may by captured with such a coarse zoning. This is especially true for the inner zones, for which the peak temperatures change more rapidly when moving from one zone to the next. Moreover, a coarse grid tends to overemphasize certain zones when a finer grid would actually find a significant change in abundance in several zones. In the context of uncertainties, this implies that certain reaction flows obtain more weight than they actually should, which results in a tendency to overestimate the uncertainty due to the lack of alternative paths.

A further problem can be identified comparing Figs.\ \ref{fig:peakhashi} and \ref{fig:peakK25}: The considered zones may not contain all zones necessary to completely follow the synthesis of light $p$ nuclei, as the zone cutoff lies too far out and several zones further inside the star may still contribute. This not only impacts the resulting uncertainties but also directly affects the final abundances. It is clearly seen, on the other hand, that a sufficient number of zones from the KEPLER model was used to assure the inclusion of all nucleosynthesis zones.

The two issues described above explain the main differences found in a comparison of the final abundance uncertainties resulting from the two codes. The coarser zoning especially affects the nuclei that do not show a focussed production in one or two zones but whose production is rather spread over many zones, such as $^{113}$In, $^{115}$Sn, $^{156,158}$Dy, $^{162}$Er, $^{168}$Yb, $^{174}$Hf, $^{184}$Os, and $^{190}$Pt. The production of the lightest $p$ nuclei is also affected by this and in addition by the inner cutoff of the zones which seems to lie too far out in the star. This is important to note, since the same set of trajectories from \citet{hashi} has been, and still is, frequently used for postprocessing studies.

\subsection{Final production uncertainties}
\label{sec:finalabuns}

Tables \ref{tab:hashiuncert} and \ref{tab:kepleruncert} provide the numerical values for the production uncertainty for each of the classical $p$ nuclides. Due to the mentioned limitations of the \citet{hashi} zones, in the following we focus on discussing the results for the 15 and 25 $M_\odot$ models obtained with the higher resolution of the KEPLER code. It has to be noted that the uncertainties shown in Figs.\ \ref{fig:hashi25}--\ref{fig:kepler15} and in tables \ref{tab:hashiuncert}, \ref{tab:kepleruncert} have been combined from the contributions of all considered zones, applying the weight given in equation (\ref{eq:normprod}) instead of a simple zone mass weighting. This is to avoid an artificial dilution and diminution of the uncertainty originating from nuclear reactions by contributions from zones where no change in the abundance of the nuclide in question occurs.

It can be seen that most uncertainties are well below a factor of two, with some exceptions both for light and heavy $p$ nuclei. Comparing this to Figs.\ \ref{fig:peakK25} and \ref{fig:peakK15}, it becomes apparent that the largest uncertainties are found for nuclides which are not mainly produced in a single zone but for which the production is rather spread out across several, sometimes many, zones. Under such varying conditions more reactions can contribute and their uncertainties accumulate.

In the interpretation of the uncertainty patterns in the context of the production of $p$ nuclei during Galactic Chemical Evolution, it has to be noted that the 15 $M_\odot$ model will contribute more to the Galactic total than the 25 $M_\odot$ model because of the dominance of 15 $M_\odot$ stars in the initial stellar mass function.

\subsection{Radiogenic isotopes}
\label{sec:radio}

As already mentioned in the introduction, knowledge of the production of the unstable nuclides $^{92}$Nb, $^{97,98}$Tc, and $^{146}$Sm is important for understanding the composition of the Early Solar System which, in turn, allows to constrain Galactic Chemical Evolution models. Moreover, \citet{dauphas} have argued that the $^{92}$Mo/$^{92}$Nb production ratio also allows to rule out proton-rich nucleosynthesis sites as sources of $p$ nuclides when they do not produce sufficient amounts of $^{92}$Nb in line with the measurements.

\begin{figure}
\includegraphics[width=\columnwidth]{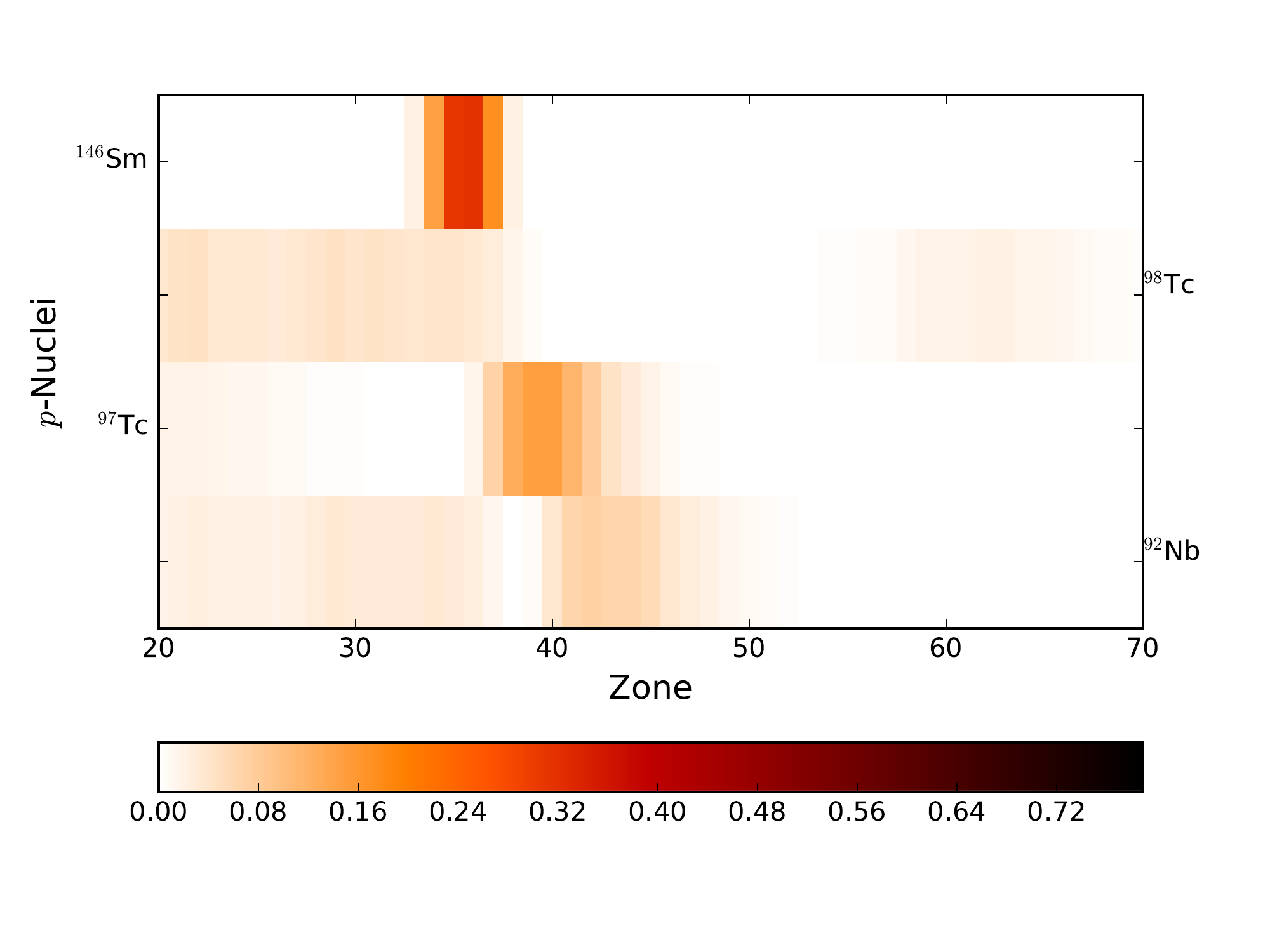}
\caption{Same as Fig.\ \ref{fig:peakK25} (25 $M_\odot$ KEPLER model) for the radiogenic nuclides. \label{fig:peakradio25}}
\end{figure}

\begin{figure}
\includegraphics[width=\columnwidth]{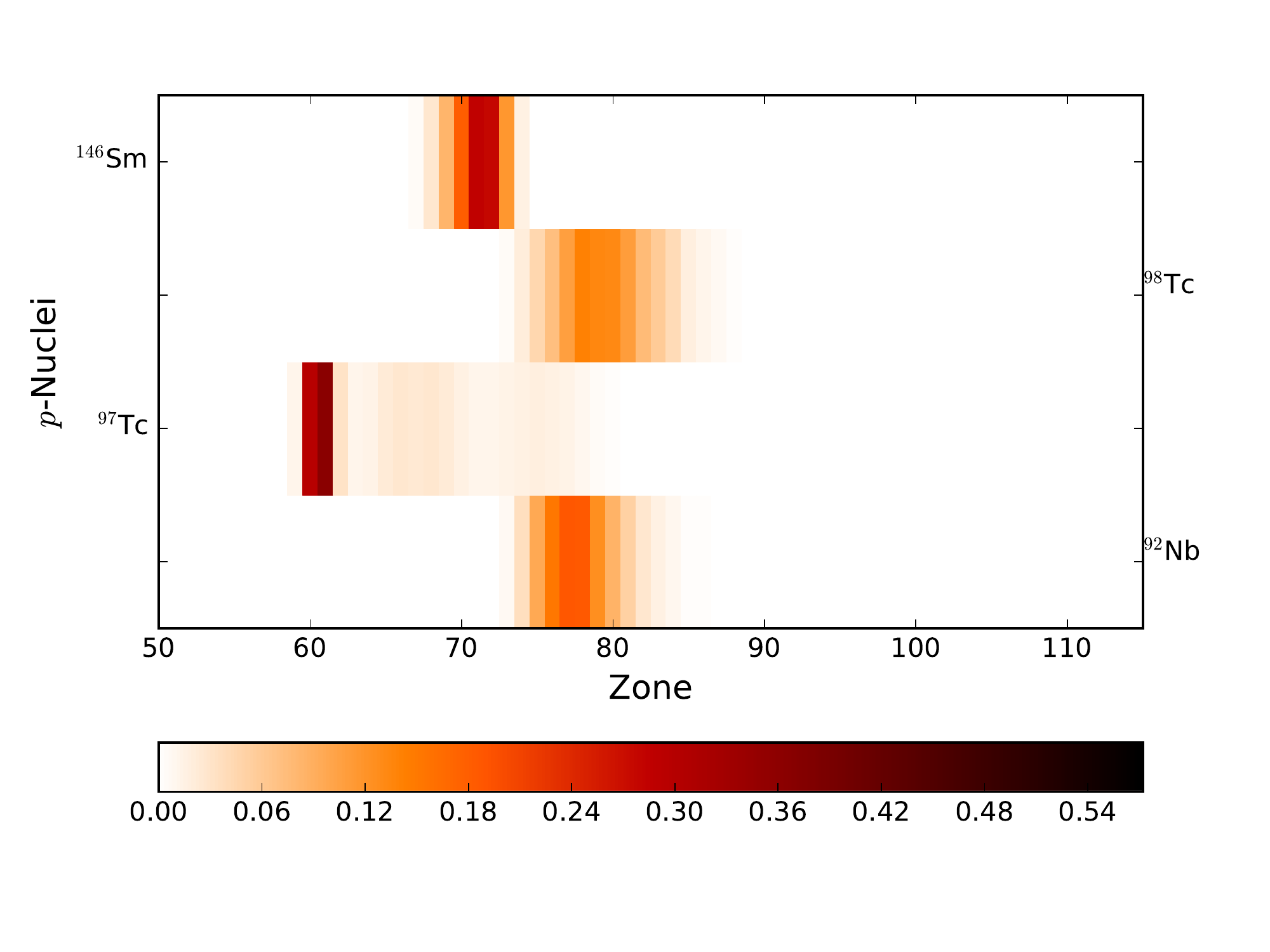}
\caption{Same as Fig.\ \ref{fig:peakradio25} but for the  15 $M_\odot$ KEPLER model. \label{fig:peakradio15}}
\end{figure}

\begin{figure}
\includegraphics[width=\columnwidth]{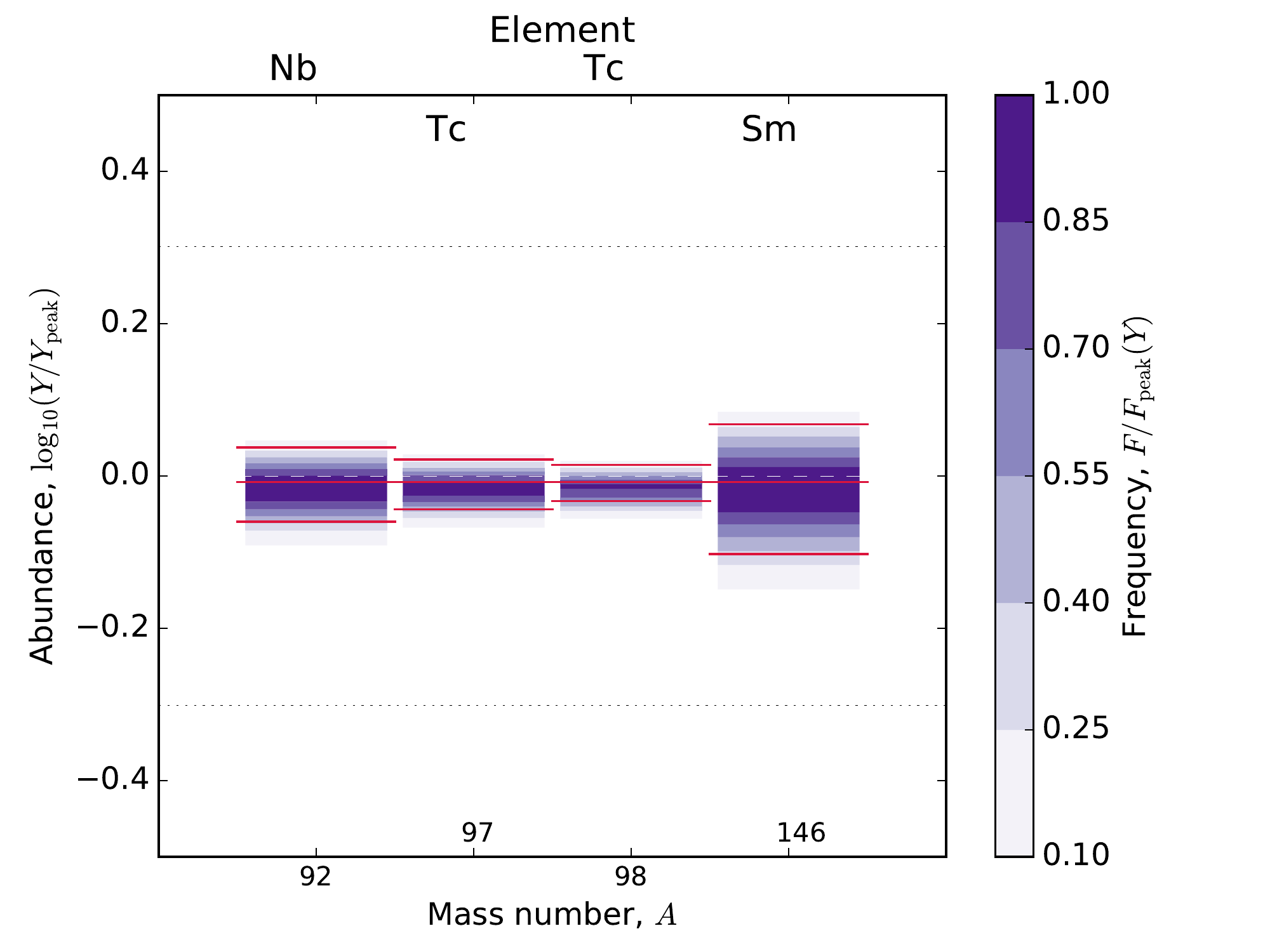}
\caption{Same as Fig.\ \ref{fig:kepler25} (25 $M_\odot$ KEPLER model) for the radiogenic nuclides. \label{fig:keplerradio25}}
\end{figure}

\begin{figure}
\includegraphics[width=\columnwidth]{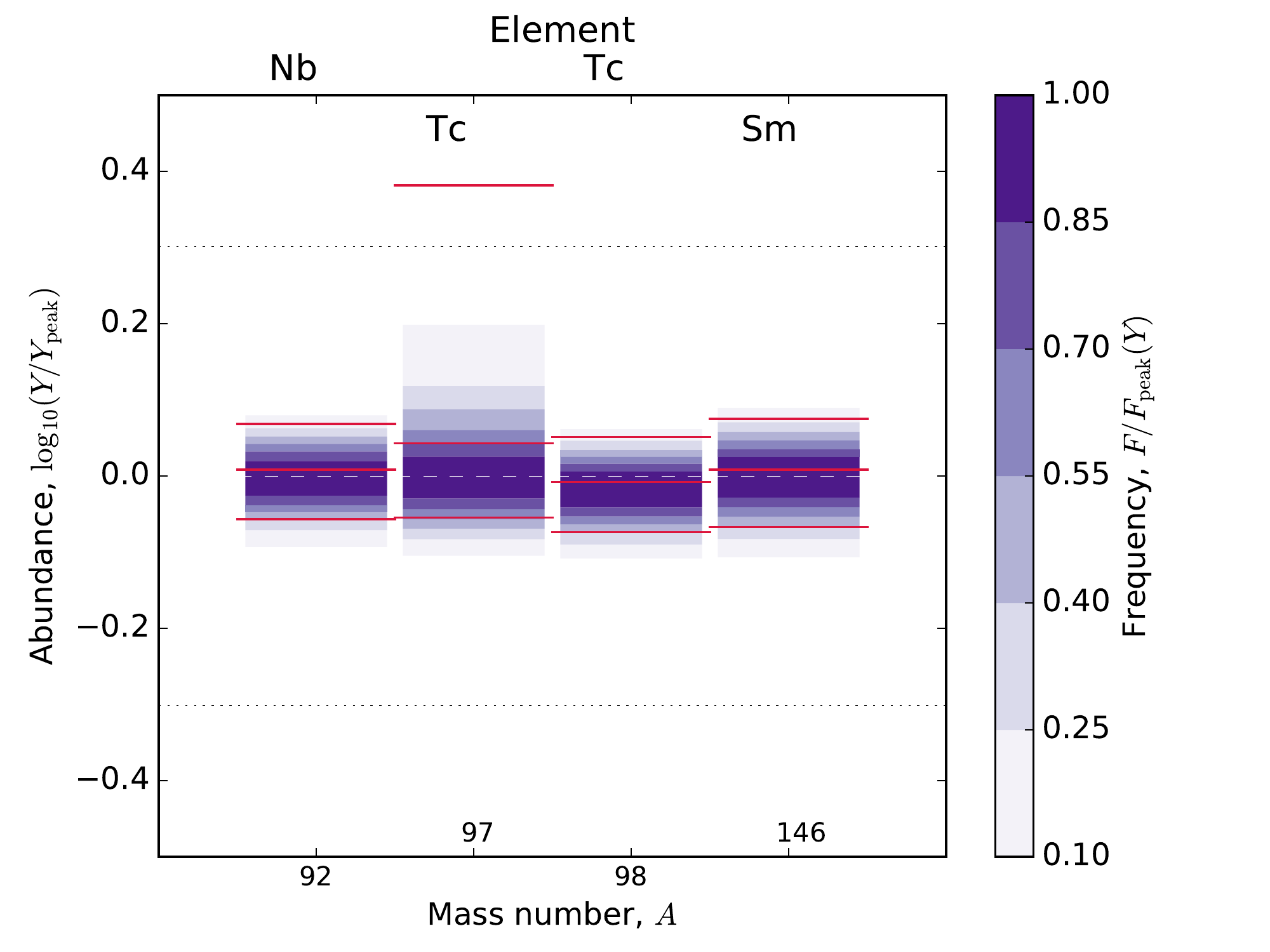}
\caption{Same as Fig.\ \ref{fig:keplerradio25} but for the  15 $M_\odot$ KEPLER model. \label{fig:keplerradio15}}
\end{figure}

Here we provide the same information for the nuclides $^{92}$Nb, $^{97,98}$Tc, and $^{146}$Sm as for the classical $p$ nuclides. Figures \ref{fig:peakradio25} and \ref{fig:peakradio15} show the zonal production for the 25 and 15 $M_\odot$ KEPLER models, respectively, and Figs.\ \ref{fig:keplerradio25} and \ref{fig:keplerradio15} the resulting production uncertainties. As before, the numerical 5\% and 95\% limits of the abundance distributions obtained in the MC procedure are given in table \ref{tab:radiouncert} for the two KEPLER models.

\begin{table}
\centering
        \caption{Same as Table \ref{tab:kepleruncert} but for the radiogenic nuclei in the 15 and 25 $M_\odot$ KEPLER models.}
        \label{tab:radiouncert}
        \begin{tabular}{lrrrr}
                \hline
                &\multicolumn{2}{c}{15 $M_\odot$ model} & \multicolumn{2}{c}{25 $M_\odot$ model} \\
                Nuclide &  up & down &  up & down\\
                \hline
$^{92}{\rm Nb}$&1.170&0.878   &1.091&0.871    \\
$^{97}{\rm Tc}$&2.408&0.882   &1.051&0.904    \\
$^{98}{\rm Tc}$&1.125&0.844   &1.034&0.927    \\
$^{146}{\rm Sm}$&1.188&0.857  &1.169&0.790   \\
\hline
\end{tabular}
\end{table}

\subsection{Key rates}
\label{sec:key}

Tables \ref{tab:key25} and \ref{tab:key15} show the key rates identified for each nuclide using the procedure described in Section \ref{sec:correlations}. These tables also include key rates for the radioactive isotopes. As mentioned above, we distinguish between level 1, 2, and 3 key rates, with level 1 being the most important ones. We consider the level 1 key rates to be our main, robust result; further key rates only become important after the uncertainty in these rates has been reduced and their relevance may also be impacted by the actual, newly determined values of the rates found on previous levels. Also shown in the tables are the obtained correlation factors on which the selection of key rates is based, for the initial run ($r_\mathrm{corr,0}$), and for the subsequent runs not varying level 1 ($r_\mathrm{corr,1}$) rates or level 2  ($r_\mathrm{corr,2}$) rates. Not all $p$ nuclides appear in the tables because only those for which a level 1, 2, or 3 key rate was found are listed. The production uncertainties for the nuclides showing in Figs.\ \ref{fig:kepler25} and \ref{fig:kepler15}, but not listed in tables \ref{tab:key25} and \ref{tab:key15}, are not dominated by a single rate but rather stem from a combination of uncertainties of several or many rates.

It has to be emphasized again that rates assigned a level 1 importance by themselves mainly determine the production \textit{uncertainty} for a specific nuclide based on the input uncertainties we chose, regardless of whether this uncertainty is large or small. If one wishes to select rates from Tables \ref{tab:key25} and \ref{tab:key15} for further theoretical or experimental study, the amount of expected reduction in the production uncertainty should be considered. The uncertainties can be found in table \ref{tab:kepleruncert}. A time-consuming, expensive measurement may not be worthwhile when the original uncertainty is already low.

Another very important point to consider when selecting possible targets for experiments is the contribution of thermally excited states to the stellar rate, as introduced in Section \ref{sec:errdefs}. The g.s.\ contribution $X_0$ to the stellar rate, as defined in equation (\ref{eq:xfactor}), is also quoted in tables \ref{tab:key25} and \ref{tab:key15} for convenience. Shown are the g.s.\ contributions for the reaction direction for which they are the largest. For capture $\leftrightarrow$ photodisintegration, this is always the capture direction whereas $X_0$ is smaller by several orders of magnitude for photodisintegration \citep{sensi}. Especially in a high-temperature nucleosynthesis process, such as the $\gamma$ process, most stellar rates are dominated by transitions from excited states and a measurement of the reaction cross section of the g.s.\ of a nucleus will not be able to provide a better constraint as the uncertainty will be mainly due to the theory error. Level 1 key rates with large $X_0$ are few, only rates relevant to the abundances of $^{92,94}$Mo, $^{96}$Ru, $^{138}$Ce, $^{144}$Sm have more than 80\% g.s.\ contribution to the stellar rate. The key rates on all levels for the radioactive nuclides also exhibit large g.s.\ contributions.

Tables \ref{tab:key25} and \ref{tab:key15} also give level 2 and 3 key rates, following the prescription of Section \ref{sec:correlations}. It is important to remember that these become ``key'' only after the uncertainties in higher level key rates have been reduced. A reduction in uncertainty in a lower-level key rate does not have much impact when higher-level rates have not been improved. Furthermore, none of the level 2 and 3 rates show $X_0\geq 0.8$, except for $^{193}$Ce + n $\leftrightarrow$ $\gamma$ + $^{194}$Ce,  $^{148}$Sm + $\alpha$ $\leftrightarrow$ $\gamma$ + $^{152}$Sm in the 25 $M_\odot$ model.

\begin{table*}
        \centering
        \caption{Key rates determining the production uncertainties in the 25 $M_\odot$ KEPLER model; shown are level 1, 2, and 3 key rates as defined in Section \ref{sec:correlations}, along with their weighted correlation factors $r_\mathrm{corr,0}$, $r_\mathrm{corr,1}$, $r_\mathrm{corr,2}$, respectively. Not all $p$ nuclides are listed but only those for which key rates were found. The lower part of the table additionally shows rates important for the production of selected unstable nuclides. Also shown for each rate are the g.s.\ contributions of the capture reaction to the stellar rate at two plasma temperatures. See text for further details.}
        \label{tab:key25}
        \begin{tabular}{crrrcccrr}
                \hline
                Nuclide & $r_\mathrm{corr,0}$ & $r_\mathrm{corr,1}$ & $r_\mathrm{corr,2}$ & Key rate & Key rate & Key rate & $X_0$ (2 GK)& $X_0$ (3 GK) \\
                &&&& Level 1 & Level 2 & Level 3 & \multicolumn{1}{c}{capture} & \multicolumn{1}{c}{capture} \\
                \hline
                $^{78}$Kr & $-0.84$ &&& $^{77}$Br + p $\leftrightarrow$ $\gamma$ + $^{78}$Kr &  && $9.63\times 10^{-2}$ &$4.44\times 10^{-2}$ \\
                 & 0.34&      0.87 &&  & $^{79}$Kr + n $\leftrightarrow$ $\gamma$ + $^{80}$Kr && $1.28\times 10^{-1}$& $7.94\times 10^{-2}$\\
                $^{92}$Mo &     $-0.74$ &&& $^{91}$Nb + p $\leftrightarrow$ $\gamma$ + $^{92}$Mo&&  & $8.88\times 10^{-1}$& $8.24\times 10^{-1}$\\
                $^{96}$Ru &      $-0.73$ &&&  $^{92}$Mo + $\alpha$ $\leftrightarrow$ $\gamma$ + $^{96}$Ru &&& 1.00&$9.86\times 10^{-1}$\\
                 & $-0.43$&     $-0.69$ &&  & $^{95}$Tc + p $\leftrightarrow$ $\gamma$ + $^{96}$Ru && $7.64\times 10^{-1}$& $6.60\times 10^{-1}$\\
                $^{102}$Pd &      $-0.87$ &&&  $^{101}$Pd + n $\leftrightarrow$ $\gamma$ + $^{102}$Pd &&& $5.62\times 10^{-1}$&$3.97\times 10^{-1}$\\
                $^{112}$Sn &     $-0.88$ &&&  $^{111}$Sn + n $\leftrightarrow$ $\gamma$ + $^{112}$Sn &&& $7.79\times 10^{-1}$&$6.73\times 10^{-1}$\\
                $^{114}$Sn &     $-0.77$ &&&  $^{113}$Sn + n $\leftrightarrow$ $\gamma$ + $^{114}$Sn &&& $1.82\times 10^{-1}$&$1.28\times 10^{-1}$\\
                $^{120}$Te & $-0.64$ & $-0.66$ &&& $^{119}$Te + n $\leftrightarrow$ $\gamma$ + $^{120}$Te &&$2.43\times 10^{-1}$&$1.77\times 10^{-1}$ \\
                $^{124}$Xe &     $-0.74$ &&&  $^{123}$Xe + n $\leftrightarrow$ $\gamma$ + $^{124}$Xe &&& $8.25\times 10^{-2}$&$4.38\times 10^{-2}$\\
                $^{126}$Xe &     $-0.75$ &&&  $^{125}$Cs + p $\leftrightarrow$ $\gamma$ + $^{126}$Ba &&& $1.17\times 10^{-1}$&$7.41\times 10^{-2}$\\
                 & 0.30&      0.64 & 0.65 &&  & $^{127}$Ba + n $\leftrightarrow$ $\gamma$ + $^{128}$Ba & $5.78\times 10^{-2}$&$3.59\times 10^{-2}$\\
                $^{130}$Ba & $-0.66$&&& $^{129}$Ba + n $\leftrightarrow$ $\gamma$ + $^{130}$Ba &&& $5.77\times 10^{-2}$&$3.55\times 10^{-2}$ \\
                $^{132}$Ba & $-0.77$&&& $^{131}$Ba + n $\leftrightarrow$ $\gamma$ + $^{132}$Ba &&& $1.07\times 10^{-1}$&$5.85\times 10^{-2}$ \\
                $^{136}$Ce & $-0.69$&&& $^{135}$Ce + n $\leftrightarrow$ $\gamma$ + $^{136}$Ce &&& $1.86\times 10^{-1}$&$8.94\times 10^{-2}$\\
                           & 0.31 & 0.72 & & & $^{139}$Ce + n $\leftrightarrow$ $\gamma$ + $^{140}$Ce &&$8.56\times 10^{-1}$&$6.09\times 10^{-1}$ \\
                $^{138}$Ce & $-0.66$&&& $^{137}$Ce + n $\leftrightarrow$ $\gamma$ + $^{138}$Ce &&& $4.16\times 10^{-1}$&$2.54\times 10^{-1}$\\
                           & $-0.16$ & $-0.19$ & $-0.66$ &&& $^{136}$Ce + n $\leftrightarrow$ $\gamma$ + $^{137}$Ce &$7.57\times 10^{-1}$&$4.70\times 10^{-1}$ \\
                $^{144}$Sm &      0.70 &&&  $^{145}$Eu + p $\leftrightarrow$ $\gamma$ + $^{146}$Gd &&& $8.06\times 10^{-1}$&$6.02\times 10^{-1}$\\
                $^{152}$Gd &     $-0.74$ &&&  $^{151}$Gd + n $\leftrightarrow$ $\gamma$ + $^{152}$Gd &&& $6.18\times 10^{-1}$&$3.87\times 10^{-1}$\\
                & 0.43 & 0.76 &&& $^{153}$Gd + n $\leftrightarrow$ $\gamma$ + $^{154}$Gd&&$5.38\times 10^{-2}$&$2.78\times 10^{-2}$\\
                & -0.14 & -0.26 & -0.73 &  &  & $^{148}$Sm + $\alpha$ $\leftrightarrow$ $\gamma$ + $^{152}$Gd &$8.14\times 10^{-1}$&$5.22\times 10^{-1}$\\
                $^{164}$Er &     $-0.78$ &&&  $^{160}$Er + $\alpha$ $\leftrightarrow$ $\gamma$ + $^{164}$Yb &&& $2.13\times 10^{-1}$&$1.24\times 10^{-1}$\\
                $^{180}$W &     $-0.83$ &&&  $^{176}$W + $\alpha$ $\leftrightarrow$ $\gamma$ + $^{180}$Os &&& $1.83\times 10^{-1}$&$1.04\times 10^{-1}$\\
                & -0.19 & -0.60 & -0.68 &  &  & $^{179}$Os + n $\leftrightarrow$ $\gamma$ + $^{180}$Os &$4.89\times 10^{-2}$
&$2.49\times 10^{-2}$\\
                $^{196}$Hg &     $-0.83$ &&&  $^{195}$Pb + n $\leftrightarrow$ $\gamma$ + $^{196}$Pb &&& $2.97\times 10^{-1}$&$1.89\times 10^{-1}$\\
                 & 0.31 &      0.70 && & $^{197}$Pb + n $\leftrightarrow$ $\gamma$ + $^{198}$Pb && $3.28\times 10^{-1}$&$2.39\times 10^{-1}$\\
                 & 0.17 & 0.35 &       0.67& && $^{199}$Pb + n $\leftrightarrow$ $\gamma$ + $^{200}$Pb & $6.37\times 10^{-1}$&$3.47\times 10^{-1}$\\
                                \hline
               $^{92}$Nb & 0.76 &&& $^{90}$Zr + p $\leftrightarrow$ $\gamma$ + $^{91}$Nb &&& 1.00&$9.95\times 10^{-1}$ \\
               $^{146}$Sm & -0.57 & -0.75 &       &  & $^{144}$Sm + $\alpha$ $\leftrightarrow$ $\gamma$ + $^{148}$Gd &&$9.99\times 10^{-1}$&$9.65\times 10^{-1}$   \\
     &  0.34 &  0.44 &  0.79 &  &  & $^{147}$Gd + n $\leftrightarrow$ $\gamma$ + $^{148}$Gd &$9.92\times 10^{-1}$&$9.28\times 10^{-1}$\\
               \hline
        \end{tabular}
\end{table*}

\begin{table*}
        \centering
        \caption{Key rates determining production uncertainties in the 15 $M_\odot$ KEPLER model; shown are level 1, 2, and 3 key rates as defined in Section \ref{sec:correlations}, along with their weighted correlation factors $r_\mathrm{corr,0}$, $r_\mathrm{corr,1}$, $r_\mathrm{corr,2}$, respectively. Not all $p$ nuclides are listed but only those for which key rates were found. The key rates shown for $^{138}$La and $^{180}$Ta do not include reactions from the $\nu$-process. The lower part of the table additionally shows rates important for the production of selected unstable nuclides. Also shown for each rate are the g.s.\ contributions of the capture reaction to the stellar rate at two plasma temperatures. See text for further details.}
        \label{tab:key15}
        \begin{tabular}{crrrcccrr}
                \hline
                Nuclide & $r_\mathrm{corr,0}$ & $r_\mathrm{corr,1}$  & $r_\mathrm{corr,2}$ & Key rate & Key rate & Key rate & $X_0$ (2 GK)& $X_0$ (3 GK) \\
                &&&& Level 1 & Level 2 & Level 3 & \multicolumn{1}{c}{capture} & \multicolumn{1}{c}{capture} \\
                \hline
                $^{78}$Kr & $-0.77$ &&&  $^{77}$Br + p $\leftrightarrow$ $\gamma$ + $^{78}$Kr &&& $9.63\times 10^{-2}$ &$4.44\times 10^{-2}$ \\
                  & $0.38$&      0.66 & && $^{79}$Kr + n $\leftrightarrow$ $\gamma$ + $^{80}$Kr && $1.28\times 10^{-1}$& $7.94\times 10^{-2}$\\
                $^{92}$Mo &     $-0.87$ &&&  $^{91}$Nb + p $\leftrightarrow$ $\gamma$ + $^{92}$Mo &&& $8.88\times 10^{-1}$& $8.24\times 10^{-1}$\\
                $^{94}$Mo &      0.78 &&&  $^{95}$Mo + n $\leftrightarrow$ $\gamma$ + $^{96}$Mo &&& $9.14\times 10^{-1}$&$7.69\times 10^{-1}$\\
                $^{96}$Ru &     $-0.67$ &&&  $^{92}$Mo + $\alpha$ $\leftrightarrow$ $\gamma$ + $^{96}$Ru &&& 1.00&$9.86\times 10^{-1}$\\
                $^{102}$Pd &     $-0.71$ &&&  $^{101}$Pd + n $\leftrightarrow$ $\gamma$ + $^{102}$Pd &&& $5.62\times 10^{-1}$&$3.97\times 10^{-1}$\\
                $^{112}$Sn &     $-0.74$ &&&  $^{111}$Sn + n $\leftrightarrow$ $\gamma$ + $^{112}$Sn &&& $7.79\times 10^{-1}$&$6.73\times 10^{-1}$\\
                $^{136}$Ce & 0.53 &      0.66 & && $^{137}$Ce + n $\leftrightarrow$ $\gamma$ + $^{138}$Ce && $4.16\times 10^{-1}$&$2.54\times 10^{-1}$\\
                $^{138}$Ce &      0.71 &&&  $^{139}$Ce + n $\leftrightarrow$ $\gamma$ + $^{140}$Ce &&& $8.71\times 10^{-1}$&$6.43\times 10^{-1}$\\
                $^{138}$La &      0.94 &&&  $^{138}$La + n $\leftrightarrow$ $\gamma$ + $^{139}$La &&& $6.18\times 10^{-1}$&$4.92\times 10^{-1}$\\
                $^{144}$Sm &      0.79 &&&  $^{145}$Eu + p $\leftrightarrow$ $\gamma$ + $^{146}$Gd &&& $8.06\times 10^{-1}$&$6.02\times 10^{-1}$\\
                $^{164}$Er &     $-0.76$ &&&  $^{160}$Er + $\alpha$ $\leftrightarrow$ $\gamma$ + $^{164}$Yb &  && $2.13\times 10^{-1}$&$1.24\times 10^{-1}$\\
                $^{168}$Yb &     $-0.80$ &&&  $^{164}$Yb + $\alpha$ $\leftrightarrow$ $\gamma$ + $^{168}$Hf &&& $2.12\times 10^{-1}$&$1.26\times 10^{-1}$\\
                 & $-0.14$&     $-0.67$ & && $^{166}$Yb + $\alpha$ $\leftrightarrow$ $\gamma$ + $^{170}$Hf && $1.80\times 10^{-1}$&$1.10\times 10^{-1}$\\
                $^{180}$Ta &     $-0.88$ &&&  $^{180}$Ta + n $\leftrightarrow$ $\gamma$ + $^{181}$Ta &&& $7.09\times 10^{-2}$&$3.96\times 10^{-2}$\\
                 & 0.09&      0.90 & && $^{179}$Ta + n $\leftrightarrow$ $\gamma$ + $^{180}$Ta && $2.37\times 10^{-1}$&$1.46\times 10^{-1}$\\
                $^{180}$W &    $-0.82$ &&&  $^{176}$W + $\alpha$ $\leftrightarrow$ $\gamma$ + $^{180}$Os &&& $1.83\times 10^{-1}$&$1.04\times 10^{-1}$\\
                $^{190}$Pt &     $-0.79$ &&&  $^{190}$Pt + n $\leftrightarrow$ $\gamma$ + $^{191}$Pt &&& $3.58\times 10^{-1}$&$1.58\times 10^{-1}$\\
                $^{196}$Hg &     $-0.86$ &&&  $^{195}$Pb + n $\leftrightarrow$ $\gamma$ + $^{196}$Pb &&& $2.97\times 10^{-1}$&$1.89\times 10^{-1}$\\
                &  0.17 &  0.64 &  0.65 &  &  & $^{197}$Pb + n $\leftrightarrow$ $\gamma$ + $^{198}$Pb &$3.28\times 10^{-1}$&$2.39\times 10^{-1}$ \\
                \hline
                $^{92}$Nb &     0.75 &&&  $^{92}$Zr + p $\leftrightarrow$ $\gamma$ + $^{93}$Nb &&& $9.91\times 10^{-1}$&$9.76\times 10^{-1}$ \\
                $^{98}$Tc &     0.89 &&&  $^{96}$Mo + p $\leftrightarrow$ $\gamma$ + $^{97}$Tc &&& $9.50\times 10^{-1}$&$8.56\times 10^{-1}$ \\
                $^{146}$Sm &   $-0.65$ &&&  $^{144}$Sm + $\alpha$ $\leftrightarrow$ $\gamma$ + $^{148}$Gd &&& $9.99\times 10^{-1}$&$9.65\times 10^{-1}$ \\
                &  0.33 &  0.79 &       &  & $^{147}$Gd + n $\leftrightarrow$ $\gamma$ + $^{148}$Gd& &$9.92\times 10^{-1}$&$9.28\times 10^{-1}$  \\
                \hline
        \end{tabular}
\end{table*}

\begin{figure}
\includegraphics[width=\columnwidth]{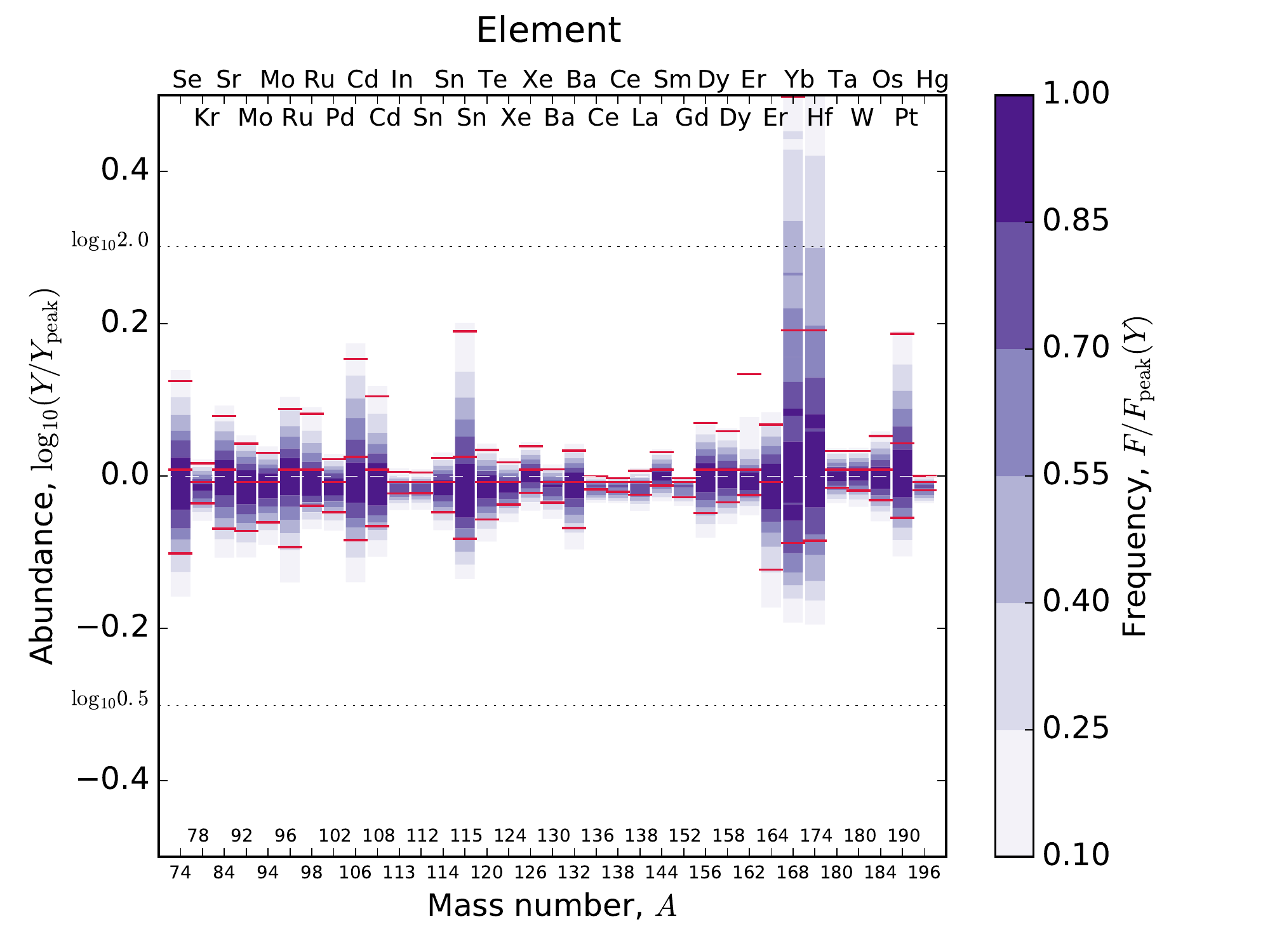}
\caption{Same as Fig.\ \ref{fig:kepler25} (25 $M_\odot$ KEPLER model) but without variation of all key rates (levels $1-3$) shown in Table \ref{tab:key25}.}
\label{fig:keyfixed25}
\end{figure}

\begin{figure}
\includegraphics[width=\columnwidth]{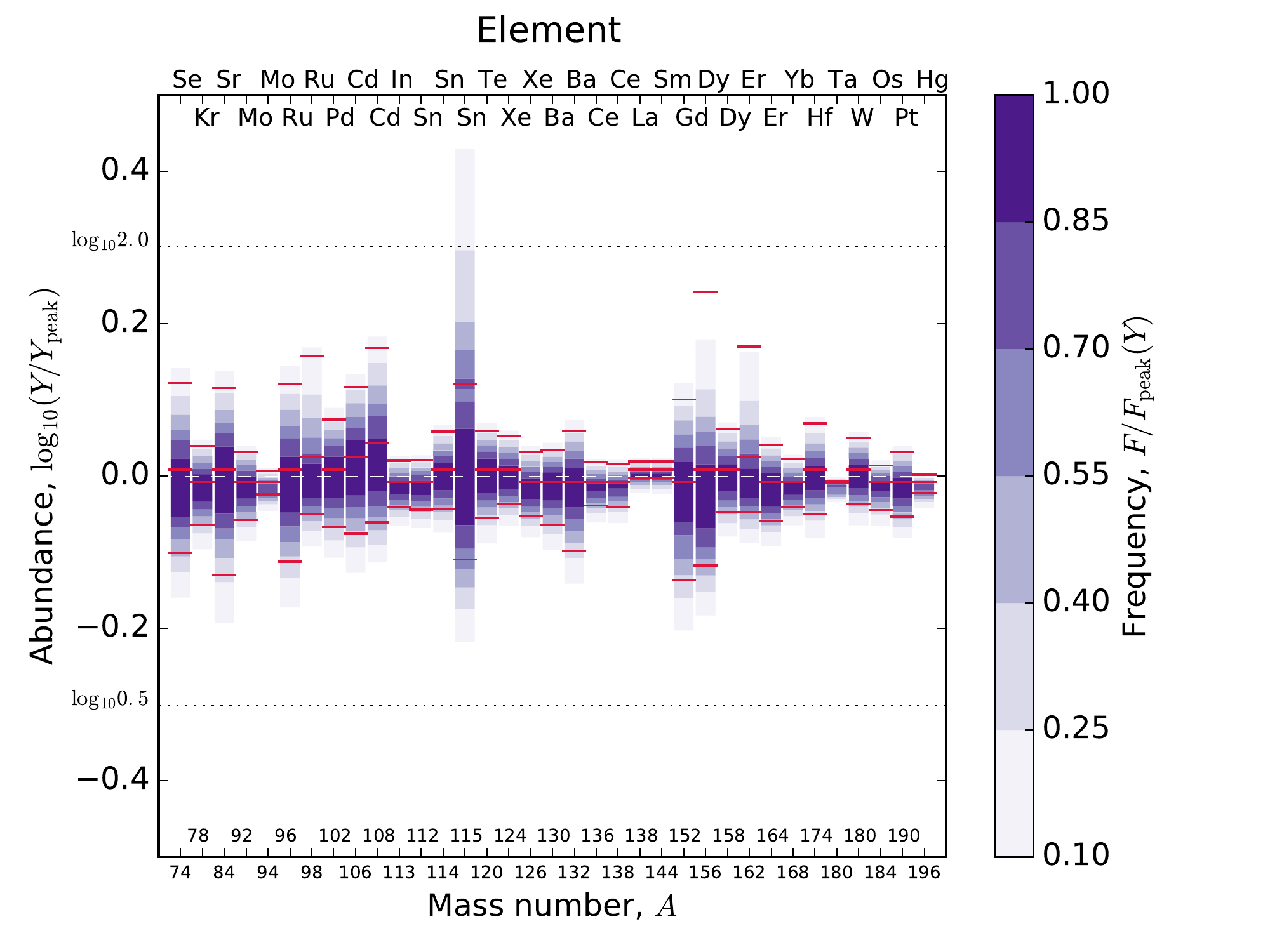}
\caption{Same as Fig.\ \ref{fig:kepler15} (15 $M_\odot$ KEPLER model) but without variation of all key rates (levels $1-2$) shown in Table \ref{tab:key15}.}
\label{fig:keyfixed15}
\end{figure}

To demonstrate the impact of a complete removal of the uncertainties in the listed key rates, Figs.\ \ref{fig:keyfixed25} and \ref{fig:keyfixed15} show the final abundance uncertainties obtained when the level $1-3$ key rates have been taken out of the variation.
Most abundance uncertainties have been reduced to a low level, with the exceptions of those of $^{115}$Sn in the 15 $M_\odot$ model and of $^{168}$Yb, $^{174}$Hf in the 25 $M_\odot$ model. It was not possible to provide key rates for these isotopes because no rate exhibited a correlation value above our threshold. This can be understood by the fact that these are isotopes whose production is not confined to a single zone but rather spread out over many zones. Thus, many different rates can contribute. Moreover, their production is feeble, anyway, and $^{115}$Sn is expected not to be a pure $p$ nuclide but receives strong contributions from the $s$- and $r$ processes (see Section \ref{sec:intro}). Although not key rates by our definition, we expect the rates $^{114}$Sn + n $\leftrightarrow$ $\gamma$ + $^{115}$Sn ($r_\mathrm{corr}=-0.59$), $^{164}$Yb + $\alpha$ $\leftrightarrow$ $\gamma$ + $^{168}$Hf ($r_\mathrm{corr}=-0.64$), and $^{170}$Hf + $\alpha$ $\leftrightarrow$ $\gamma$ + $^{174}$W ($r_\mathrm{corr}=-0.42$), to contribute to the abundance uncertainties of $^{115}$Sn, $^{168}$Yb and $^{174}$Hf, respectively.

\begin{figure}
\includegraphics[width=\columnwidth]{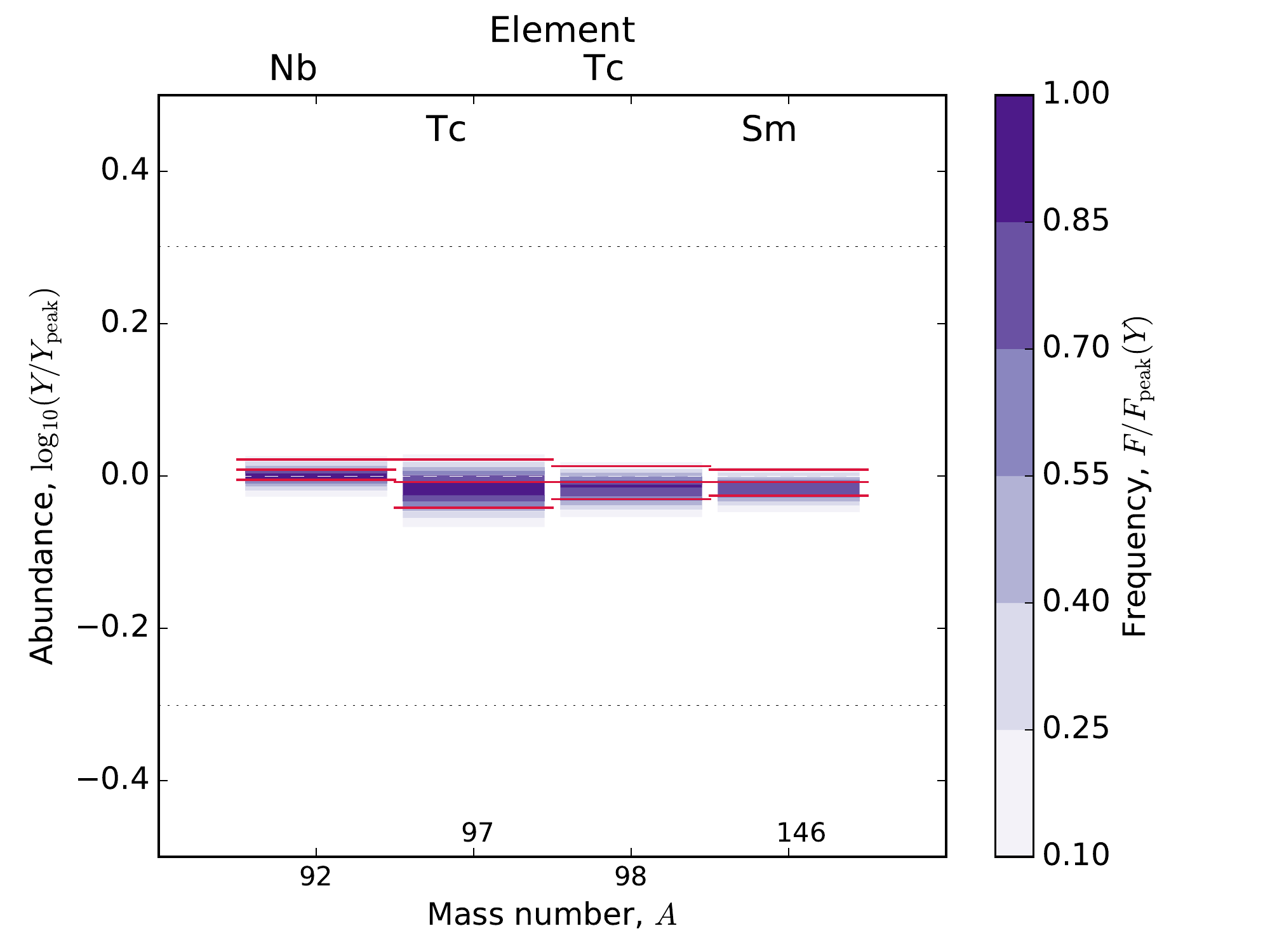}
\caption{Same as Fig.\ \ref{fig:keyfixed25} (25 $M_\odot$ KEPLER model) for the radiogenic nuclides. \label{fig:fixedradio25}}
\end{figure}

\begin{figure}
\includegraphics[width=\columnwidth]{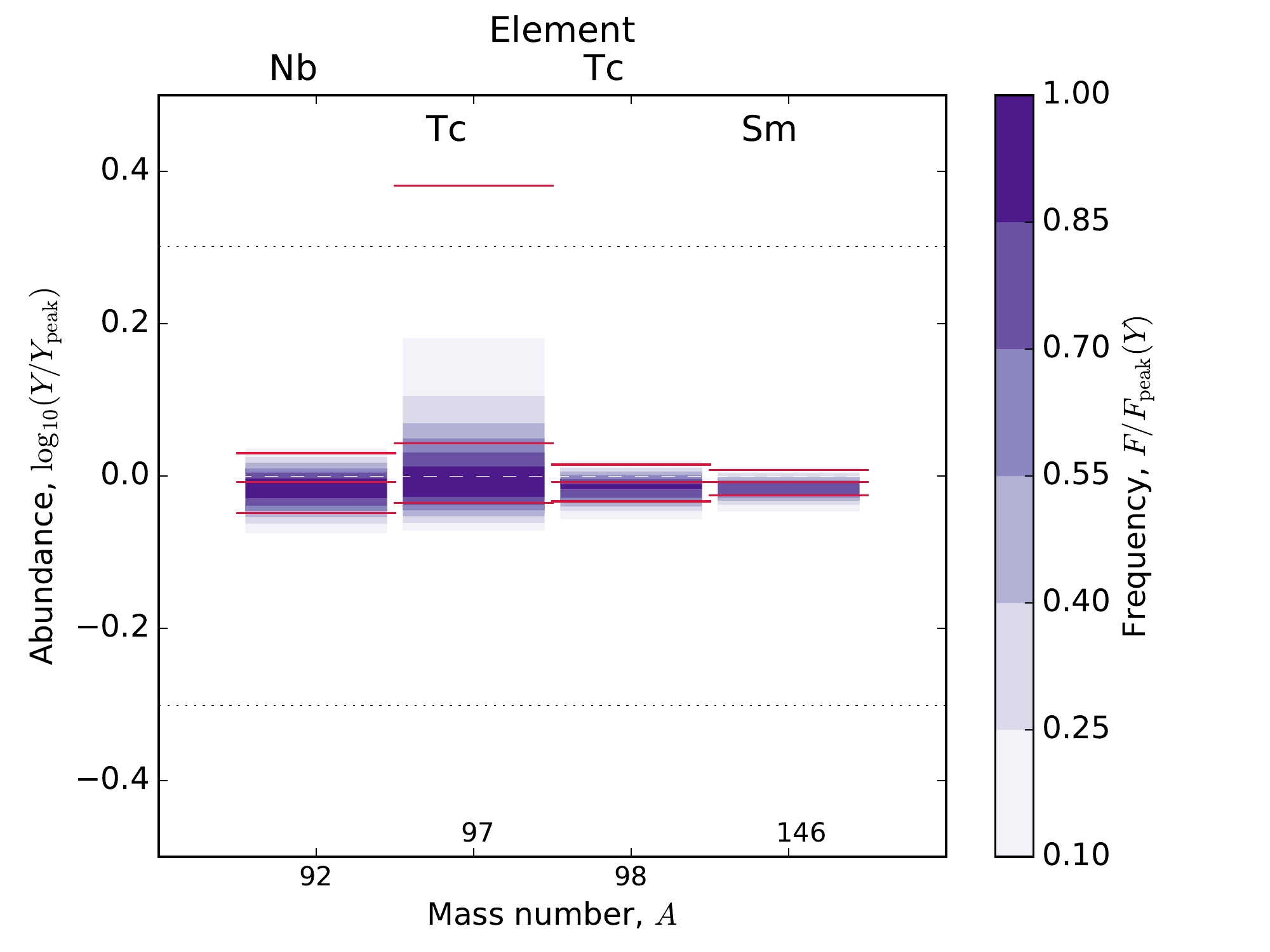}
\caption{Same as Fig.\ \ref{fig:keyfixed15} (15 $M_\odot$ KEPLER model) for the radiogenic nuclides. \label{fig:fixedradio15}}
\end{figure}

The key rates influencing the production of the radioactive nuclides $^{92}$Nb, $^{97,98}$Tc, and $^{146}$Sm are also given in the tables \ref{tab:key25} and \ref{tab:key15}. The remaining uncertainties after having taken out all key rates from the MC variation are shown in Figs.\ \ref{fig:fixedradio25} and \ref{fig:fixedradio15} for the 25 and 15 $M_\odot$ KEPLER models, respectively. With the exception of $^{97}$Tc in the 15 $M_\odot$ model, the already small uncertainties were further reduced. In the 15 $M_\odot$ model, $^{97}$Tc exhibits a strongly asymmetric uncertainty which is caused by the combined uncertainties of many reactions. Among these, while not being a key rate clearly dominating the uncertainty, $^{96}$Ru + n $\leftrightarrow$ $\gamma$ + $^{97}$Ru has the strongest correlation ($r_\mathrm{corr}=-0.384$).

Table \ref{tab:key25} does not list $^{74}$Se, $^{84}$Sr, $^{94}$Mo, $^{96,98}$Ru, $^{106,108}$Cd, $^{113}$In, $^{115}$Sn, $^{138}$La, $^{156,158}$Dy, $^{162}$Er, $^{168}$Yb, $^{174}$Hf, $^{180}$Ta, $^{184}$Os, and $^{97,98}$Tc as no key rates were found for these nuclei in the 25 $M_\odot$ model. Inspection of Figs.\ \ref{fig:keyfixed25}, \ref{fig:keyfixed15} shows that the production uncertainties are low for these nuclides, anyway. They are results of the combined uncertainties in several rates. For completeness, further rates with correlations $|r_\mathrm{corr,3}|\geq 0.4$ (and not already shown in table \ref{tab:key25}) contributing to the uncertainties of the above nuclides are listed in table \ref{tab:further25}. It has to be noted that these rates are not responsible on their own for the remaining uncertainties and a better determination may not directly lead to a further strong uncertainty reduction.

\begin{table*}
\centering
\caption{Further rates with $|r_\mathrm{corr,3}|\geq 0.4$ in the 25 $M_\odot$ KEPLER model, for nuclides without key rates in table \ref{tab:key25}. These do not determine the remaining production uncertainties on their own.\label{tab:further25}}
\begin{tabular}{rrcrr}
\hline
Nuclide & $r_\mathrm{corr,3}$ & Rate & $X_0$ (2 GK) & $X_0$ (3 GK) \\
\hline
$^{74}$Se & $-0.5$ & $^{73}$As + p $\leftrightarrow$ $\gamma$ + $^{74}$Se & $3.39\times 10^{-1}$&$2.41\times 10^{-1}$ \\
          & $-0.4$ & $^{70}$Ge + $\alpha$ $\leftrightarrow$ $\gamma$ + $^{74}$Se & $9.87\times 10^{-1}$&$9.15\times 10^{-1}$ \\
          & $-0.4$ & $^{75}$Se + n $\leftrightarrow$ $\gamma$ + $^{76}$Se &$4.37\times 10^{-1}$ & $3.22\times 10^{-1}$\\
$^{84}$Sr & $-0.6$ & $^{83}$Rb + p $\leftrightarrow$ $\gamma$ + $^{84}$Sr &$2.83\times 10^{-1}$ &$2.47\times 10^{-1}$ \\
$^{94}$Mo & 0.6    & $^{95}$Mo + n $\leftrightarrow$ $\gamma$ + $^{96}$Mo & $8.93\times 10^{-1}$& $7.59\times 10^{-1}$\\
          & $-0.4$ & $^{93}$Mo + n $\leftrightarrow$ $\gamma$ + $^{94}$Mo & $9.98\times 10^{-1}$& $9.71\times 10^{-1}$\\
$^{96}$Ru & $-0.6$ & $^{95}$Ru + n $\leftrightarrow$ $\gamma$ + $^{96}$Ru & $9.90\times 10^{-1}$& $9.23\times 10^{-1}$\\
          & $-0.4$ & $^{105}$Cd + n $\leftrightarrow$ $\gamma$ + $^{106}$Cd & $5.25\times 10^{-1}$& $3.71\times 10^{-1}$\\
          & $-0.4$ & $^{109}$In + p $\leftrightarrow$ $\gamma$ + $^{110}$Sn & $9.89\times 10^{-1}$& $9.28\times 10^{-1}$\\
$^{98}$Ru & $-0.6$ & $^{97}$Ru + n $\leftrightarrow$ $\gamma$ + $^{98}$Ru & $8.07\times 10^{-1}$& $6.26\times 10^{-1}$\\
$^{106}$Cd & $-0.6$ & $^{105}$Cd + n $\leftrightarrow$ $\gamma$ + $^{106}$Cd & $5.25\times 10^{-1}$& $3.71\times 10^{-1}$\\
          & $0.4$ & $^{109}$In + p $\leftrightarrow$ $\gamma$ + $^{110}$Sn & $9.89\times 10^{-1}$& $9.28\times 10^{-1}$ \\
$^{108}$Cd & $-0.6$ & $^{107}$Cd + n $\leftrightarrow$ $\gamma$ + $^{108}$Cd & $6.19\times 10^{-1}$& $4.22\times 10^{-1}$\\
$^{113}$In & $0.5$ & $^{114}$In + n $\leftrightarrow$ $\gamma$ + $^{115}$In & $1.94\times 10^{-1}$& $9.60\times 10^{-2}$\\
$^{115}$Sn & $-0.4$ & $^{114}$Sn + n $\leftrightarrow$ $\gamma$ + $^{115}$Sn & $9.93\times 10^{-1}$& $9.14\times 10^{-1}$\\
$^{168}$Yb & $-0.6$ & $^{164}$Yb + $\alpha$ $\leftrightarrow$ $\gamma$ + $^{168}$Hf & $2.14\times 10^{-1}$& $1.28\times 10^{-1}$\\
$^{174}$Hf & $-0.4$ & $^{170}$Hf + $\alpha$ $\leftrightarrow$ $\gamma$ + $^{174}$W & $1.78\times 10^{-1}$& $1.08\times 10^{-1}$\\
\hline
$^{97}$Tc & $0.5$ & $^{98}$Tc + n $\leftrightarrow$ $\gamma$ + $^{99}$Tc & $2.83\times 10^{-1}$& $2.25\times 10^{-1}$\\
          & $-0.5$ & $^{96}$Tc + n $\leftrightarrow$ $\gamma$ + $^{97}$Tc & $3.00\times 10^{-1}$& $2.53\times 10^{-1}$\\
\hline
\end{tabular}
\end{table*}

\begin{table*}
\centering
\caption{Further rates with $|r_\mathrm{corr,3}|\geq 0.4$ in the 15 $M_\odot$ KEPLER model, for nuclides without key rates in table \ref{tab:key15}. These do not determine the remaining production uncertainties on their own.\label{tab:further15}}
\begin{tabular}{rrcrr}
\hline
Nuclide & $r_\mathrm{corr,3}$ & Rate & $X_0$ (2 GK) & $X_0$ (3 GK) \\
\hline
$^{74}$Se & $0.4$ & $^{75}$As + p $\leftrightarrow$ n + $^{75}$Se & $3.53\times 10^{-1}$& $1.76\times 10^{-1}$\\
          & $-0.4$ & $^{73}$As + p $\leftrightarrow$ $\gamma$ + $^{74}$Se & $3.39\times 10^{-1}$&$2.41\times 10^{-1}$ \\
$^{84}$Sr & $0.6$ & $^{84}$Sr + n $\leftrightarrow$ $\gamma$ + $^{85}$Sr & $9.31\times 10^{-1}$& $7.27\times 10^{-1}$\\
          & $-0.5$ & $^{83}$Rb + p $\leftrightarrow$ $\gamma$ + $^{84}$Sr & $2.83\times 10^{-1}$ &$2.47\times 10^{-1}$ \\
$^{98}$Ru & $-0.6$ & $^{97}$Ru + n $\leftrightarrow$ $\gamma$ + $^{98}$Ru & $8.07\times 10^{-1}$& $6.26\times 10^{-1}$ \\
$^{106}$Cd & $-0.6$ & $^{105}$Cd + n $\leftrightarrow$ $\gamma$ + $^{106}$Cd & $5.25\times 10^{-1}$& $3.71\times 10^{-1}$ \\
          & $0.6$ & $^{109}$In + p $\leftrightarrow$ $\gamma$ + $^{110}$Sn & $9.89\times 10^{-1}$& $9.28\times 10^{-1}$ \\
$^{108}$Cd & $-0.6$ & $^{107}$Cd + n $\leftrightarrow$ $\gamma$ + $^{108}$Cd & $6.19\times 10^{-1}$& $4.22\times 10^{-1}$ \\
           & 0.4    & $^{109}$In + p $\leftrightarrow$ $\gamma$ + $^{110}$Sn & $9.89\times 10^{-1}$& $9.28\times 10^{-1}$ \\
$^{113}$In & $0.6$ & $^{113}$Sn + n $\leftrightarrow$ $\gamma$ + $^{114}$Sn & $1.89\times 10^{-1}$& $1.37\times 10^{-1}$\\
$^{114}$Sn & $-0.6$ & $^{113}$Sn + n $\leftrightarrow$ $\gamma$ + $^{114}$Sn & $1.89\times 10^{-1}$& $1.37\times 10^{-1}$ \\
$^{115}$Sn & $-0.6$ & $^{114}$Sn + n $\leftrightarrow$ $\gamma$ + $^{115}$Sn & $9.93\times 10^{-1}$& $9.14\times 10^{-1}$ \\
$^{120}$Te & $0.5$ & $^{121}$Te + n $\leftrightarrow$ $\gamma$ + $^{122}$Te & $2.02\times 10^{-1}$&$9.50\times 10^{-2}$ \\
$^{124}$Xe & $-0.5$ & $^{123}$Xe + n $\leftrightarrow$ $\gamma$ + $^{124}$Xe & $8.19\times 10^{-2}$& $4.78\times 10^{-2}$\\
$^{130}$Ba & $-0.5$ & $^{130}$Ba + n $\leftrightarrow$ $\gamma$ + $^{131}$Ba & $3.75\times 10^{-1}$& $1.65\times 10^{-1}$\\
           & $0.5$ & $^{131}$Ba + n $\leftrightarrow$ $\gamma$ + $^{132}$Ba & $1.07\times 10^{-1}$& $5.85\times 10^{-2}$\\
$^{132}$Ba & $0.4$ & $^{133}$Ba + n $\leftrightarrow$ $\gamma$ + $^{134}$Ba & $1.17\times 10^{-1}$& $6.91\times 10^{-2}$\\
$^{152}$Gd & $-0.6$ & $^{152}$Gd + n $\leftrightarrow$ $\gamma$ + $^{153}$Gd & $4.39\times 10^{-1}$& $1.97\times 10^{-1}$\\
           & $0.4$ & $^{153}$Gd + n $\leftrightarrow$ $\gamma$ + $^{154}$Gd & $5.38\times 10^{-2}$& $2.78\times 10^{-2}$\\
$^{158}$Dy & $-0.6$ & $^{157}$Dy + n $\leftrightarrow$ $\gamma$ + $^{158}$Dy & $8.23\times 10^{-2}$& $4.12\times 10^{-2}$\\
           & 0.5    & $^{156}$Dy + n $\leftrightarrow$ $\gamma$ + $^{157}$Dy & $1.49\times 10^{-1}$&  $7.70\times 10^{-2}$\\
$^{162}$Er & $-0.5$ & $^{158}$Er + $\alpha$ $\leftrightarrow$ $\gamma$ + $^{162}$Yb & $3.10\times 10^{-1}$& $1.71\times 10^{-1}$\\
$^{174}$Hf & $-0.4$ & $^{174}$Hf + n $\leftrightarrow$ $\gamma$ + $^{175}$Hf & $1.01\times 10^{-1}$& $5.56\times 10^{-2}$\\
$^{184}$Os & $-0.5$ & $^{184}$Os + n $\leftrightarrow$ $\gamma$ + $^{185}$Os & $1.39\times 10^{-1}$& $7.78\times 10^{-2}$\\
$^{196}$Hg & $0.5$ & $^{199}$Pb + n $\leftrightarrow$ $\gamma$ + $^{200}$Pb & $4.21\times 10^{-1}$& $2.03\times 10^{-1}$\\
\hline
$^{97}$Tc & $-0.4$ & $^{96}$Ru + n $\leftrightarrow$ $\gamma$ + $^{97}$Ru & 1.00&$9.91\times 10^{-1}$ \\
\hline
\end{tabular}
\end{table*}

\section{Conclusions}

For the first time we have performed a comprehensive, large-scale MC study of nucleosynthesis in the $\gamma$ process in massive stars, varying reactions on targets from Fe to Bi. Temperature-dependent stellar reaction rate uncertainties were individually assigned to the reactions, allowing a quantification of the uncertainties in final $p$-nucleus production due to nuclear input. Our approach also allowed identification of a number of key rates -- which contribute most to these final uncertainties -- by a well-defined, automated method. Overall, the uncertainties were found to be modest, better than a factor of two. The remaining uncertainties are mainly due to theoretically predicted rates of reactions on unstable nuclei and of excited state contributions. Only few uncertainties could be reduced directly by measuring g.s.\ cross sections. These are the ones with large g.s.\ contributions to the stellar rate, which occur only for target nuclei in the mass range of the light $p$ nuclei.

Abundances and their uncertainties were studied in three different stellar models, two based on the same code and input but for two progenitor masses, and one for an additional, widely-used stellar model. A direct comparison revealed that the older model used too crude a mass grid to be able to follow properly the temperature evolution in the $\gamma$-process layers. The coarse grid and an inner cutoff at too large a radius led to discrepancies mainly in the prediction of light $p$ nuclei and a number of heavier $p$ nuclides whose production is spread out over many zones.

The results of our study are expected to be useful for both astrophysicists and nuclear physicists. They can be incorporated in Galactic Chemical Evolution models to assess uncertainties in Galactic $p$-nucleus production, although it would be desirable to extend this investigation to a wider range of progenitor masses and initial metallicities to fully capture the uncertainty distribution. Experimental and theoretical nuclear physicists can see which reaction rates need to be improved to successfully reduce the uncertainties.

This maiden voyage of our MC framework also worked as a proof of concept, showing that it is possible to perform large-scale rate variations in extended reaction networks and use the MC method to not only quantify combined uncertainties of all rates but also derive the key rates that contribute the most to those uncertainties. We plan to apply our method to further nucleosynthesis processes, such as the $\gamma$-process in thermonuclear supernovae (in preparation), the weak $s$ process in massive stars \citep{weakNobuya}, the main $s$ process \citep{cescutti}, and the $\nu p$-, $rp$-, and $r$-processes.

\section*{Acknowledgements}

We thank U. Frischknecht for his initial help in the development of the MC framework. This work was partially supported by the European Research Council (grants GA 321263-FISH and EU-FP7-ERC-2012-St Grant 306901) and the UK Science and Technology Facilities Council (grants ST/M000958/1, ST/M001067/1).
This work used the DIRAC Shared Memory Processing system at the University
of Cambridge, operated by the COSMOS Project at the Department of
Applied Mathematics and Theoretical Physics on behalf of the STFC
DiRAC HPC Facility (www.dirac.ac.uk).
This equipment was funded by BIS National E-infrastructure capital
grant ST/J005673/1, STFC capital grant ST/H008586/1, and STFC DiRAC
Operations grant ST/K00333X/1. DiRAC is part of the National
E-Infrastructure. The University of Edinburgh is a charitable body, registered in Scotland, with Registration No.\ SC005336.








%
%


\bsp	
\label{lastpage}
\end{document}